\documentclass[twocolumn]{aastex631}

\newcommand\virga{\texttt{Virga}}
\newcommand\eddysed{\texttt{EddySed}}
\usepackage[version=4]{mhchem}
\newcommand\picaso{\texttt{PICASO}}

\defcitealias{virgaV2batalha}{Virga V2.0 Zenodo}
\defcitealias{batalha2025virga}{Batalha et al. 2025}
\revised{September 5, 2025}
\shorttitle{Fast Fluffy Fractals for {\virga}}
\shortauthors{Moran and Lodge et al.}

\begin{document}

\title{Fractal Aggregate Aerosols in the Virga Cloud Code I: Model Description and Application to a Benchmark Cloudy Exoplanet \\ }

\author[0000-0002-6721-3284]{Sarah E. Moran}
\altaffiliation{NHFP Sagan Fellow}
\affiliation{NASA Goddard Space Flight Center, Greenbelt, MD 20771, USA}
\affiliation{Space Telescope Science Institute, Baltimore, MD 21218, USA}
\correspondingauthor{Sarah E. Moran \& Matt G. Lodge}
\email{moran.sarahe@gmail.com, m.g.lodge@bristol.ac.uk}



\author[0000-0002-9733-0617]{Matt G. Lodge}
\affiliation{School of Physics, University of Bristol
Bristol, UK}

\author[0000-0003-1240-6844]{Natasha E. Batalha}
\affiliation{NASA Ames Research Center,
Moffett Field, CA 94035, USA}

\author[0000-0003-3290-6758]{Kazumasa Ohno}
\affiliation{Division of Science, National Astronomical Observatory of Japan, Mitaka-shi, JP}

\author{Sanaz Vahidinia}
\affiliation{NASA Ames Research Center,
Moffett Field, CA 94035, USA}
\affiliation{NASA Headquarters, Washington, D.C., USA}

\author[0000-0002-5251-2943]{Mark S. Marley}
\affiliation{Department of Planetary Sciences and Lunar and Planetary Laboratory, University of Arizona, Tucson, AZ, USA}

\author[0000-0003-4328-3867]{Hannah R. Wakeford}
\affiliation{School of Physics, University of Bristol
Bristol, UK}

\author[0000-0003-4813-7922]{Zo{\"e} M. Leinhardt}
\affiliation{School of Physics, University of Bristol
Bristol, UK}




\begin{abstract}

We introduce new functionality to treat fractal aggregate aerosol particles within the {\virga} cloud modeling framework. Previously, the open source cloud modeling code {\virga} \citep{batalha2025virga}, the Python version of \texttt{EddySed} \citep{ackerman2001}, assumed spherical particles to compute particle mass and size distributions throughout the atmosphere. The initial release of {\virga} also assumed spherical particles to compute Mie scattering properties, which include the single scattering albedo, asymmetry parameter, and optical depth as a function of particle radius and composition. However, extensive evidence from Solar system aerosols, astrophysical disks and dust, and Earth climate studies suggests that non-spherical aggregate particles are common compared to idealized compact spherical particles. Following recent advances in microphysical and opacity modeling, we implement a simple parametrization for dynamical and optical (modified mean field theory) effects of fractal aggregate particles into {\virga}. We then use this new functionality to perform a case study using basic planetary parameters similar to the well-characterized, aerosol-laden mini-Neptune GJ~1214~b, using KCl clouds made of aggregate particles. We choose KCl to most directly explore comparisons to previous studies. We demonstrate 1) how our method compares to previous fractal aggregate particle treatments and 2) how our new fractal treatment affects theoretical spectra of cloudy atmospheres. Overall, our model is faster and more flexible for a wider range of parameter space than previous studies. We explore the limitations of our modeling set-up and offer guidance for future investigations using our framework.

\end{abstract}

\section{Introduction} \label{sec:intro}

Both clouds and/or hazes exist in every planetary (or substellar) atmosphere. Detections and inferences of clouds and/or hazes abound across the Solar system \citep[e.g.,][]{tomasko2010,chavez2023,Sanchez-lavega2023}, across the brown dwarf sequence \citep[e.g.,][]{Cushing2006,suarez2022,Miles2023}, and across a wide range of exoplanets \citep[e.g.,][]{kreidberg2014,Gao2020,Grant2023,Brande2024}. Steep spectral slopes \citep[e.g.,][]{ohno2020b}, muted spectral features in transmission \citep[e.g.,][]{Wakeford2019}, and high albedo in emission observations \citep[e.g.,][]{schlawin2024,Coulombe2025,Morel2025} have all been widely interpreted as evidence for the presence of aerosols in exoplanetary atmospheres. The most recent extensive review of such studies may be found in \citet{gao2021}, with further discussion in previous reviews such as \citet{Helling2019,Marley2013,Marley2015}.

As detailed more extensively in \citet{sanaz2024}, most exoplanetary cloud (and haze) modeling focuses on spherical particles for the simple reason of computational efficiency, as even using Mie calculations for particle opacity, rather than simple ``grey'' or Rayleigh slope scalings \citep[e.g.,][]{Moran2018,Macdonald2017}, is fairly recent in exoplanet retrieval codes \citep{Naskedkin2024,Mullens2024}. However, both protoplanetary disk studies and Solar system atmospheric modeling have included non-spherical and/or aggregate particle treatments for decades \citep[e.g.,][]{toon1980physical,West&Smith91,Cabane+93,
West2004,tomasko2008model,Min2006,Min2015,ohno2021haze}. Aggregate particles can form via coagulation from the nucleation of spherical or crystalline single monomer grains, and better explain numerous aspects of observational data from Solar system and Earth studies. Evidence for such particles from remote sensing includes enhanced scattering at blue wavelengths (indicative of smaller particles) simultaneous with strong forward scattering (indicative of large particles) \citep{ohno2021haze}, which is widely observed on planets and moons from Saturn \citep{West2004}, to Triton and Pluto \citep[e.g.,][]{lavvas2021}, to Titan \citep{tomasko2008model}, and throughout astrophysical disks and comets \citep[e.g.,][]{Kimura2006}. In addition, laboratory studies frequently find that monodisperse spherical particles are not the naturally occurring form of suspended aerosol \citep[e.g.,][]{He2018,hamill2024}.


A fraction of exoplanetary aerosol modeling uses the optical properties of non-spherical or non-homogeneous particles while either remaining agnostic to or prescribing spherical shapes to particles' atmospheric transport \citep{Kopparla2016,Arney2017, Min2015,Dyrek2024,exolyn2024,kiefer2024}. A smaller fraction of exoplanet-focused aerosol modeling studies have also used self-consistent aggregate particle properties coupling both the optical and dynamical effects of fractal cloud or haze particles \citep[i.e.,][]{Adams2019,ohno2020,Samra2020,Samra2022}. These prior self-consistent approaches have all involved microphysical models, which though fully physically realistic and accurate, can be computationally expensive. This computational cost can be prohibitive for atmospheric interpretation, especially when needed to vary over wide parameter space where many key planetary parameters such as atmospheric composition, mixing and transport, and thermal structure are unknown. Furthermore, the self-consistent approaches rely on the prior knowledge of microphysics affecting aerosol morphology, which risks bias against unidentified physical processes.

At the same time, inverse modeling approaches like retrievals \citep{macdonaldbatalha}, while fast and well-suited for parameter space sweeps, do not have the grounding in cloud formation physics that can often offer useful insight into atmospheric processes. In such cases, parametrized yet physically grounded models such as \texttt{EddySed} \citep{ackerman2001} have widespread usage in exoplanetary and brown dwarf literature \citep[e.g.,][]{roellig2004,Leggett2010,morley2013}. \texttt{EddySed} was updated and re-released with a beta Python version called {\virga} (Version 0.0) \citep{batalha2020,rooney2022}, which has also seen widespread use in exoplanet studies \citep[e.g.,][]{ahrer2023,Grant2023,Brande2024,Boehm2025}. {\virga} uses a mass balance approach to solve for the particle size distribution of spherical particles and then computes their opacity and scattering properties. This version and its nuances compared with the original Fortran \texttt{EddySed} are more fully described in the Version~1 release \citep{batalha2025virga}. However, given the importance of non-spherical cloud particles, {\virga} lacked some key capability to generate physically realistic cloud distributions.

To remedy this, here we include a new parametrization within {\virga} for the dynamical behavior of fractal aggregates, which expands upon the work of \citet{ohno2020}, \citet{Tazaki2021}, and the scaling relationships discussed in \citet{sanaz2024}, detailed in Section \ref{sec:dynamics}. We also incorporate Modified Mean Field Theory \citep[MMF;][]{Tazaki2018,Tazaki2021} as a way to compute the absorption, scattering, and extinction of non-spherical fractal aggregates, detailed in Section \ref{sec:optics}. In Section \ref{sec:results}, we explore how our methodology and observational simulations compare to previous microphysical studies, in particular \citet{ohno2020} and \citet{Adams2019}, as these works modeled the same general planetary conditions but with key differences in fractal particle assumptions. We discuss caveats and limitations of our approach, avenues for future studies using our aggregate-enabled {\virga} model, and present our conclusions in Section \ref{sec:conclusion}.

\section{Dynamics of Fractal Aggregate Particles} \label{sec:dynamics}


{\virga} implements the {\eddysed} model \citep{ackerman2001, GaoMarleyAckerman2018} to balance upward transport, such as convection, and sedimentation as governed by:
\begin{equation} \label{Eq:eddysed}
    -K_{zz}\frac{\delta q_t}{\delta z} - f_\mathrm{sed} w_* q_c=0,
\end{equation}

\noindent where $K_{zz}$ is the eddy-diffusion coefficient, $f_\mathrm{sed}$ is the sedimentation efficiency (the ratio of the sedimentation velocity to the convective velocity), $q_c$ is the condensate mass mixing ratio, $q_t$ is the total (vapor + condensate) mass mixing ratio, $z$ is the altitude, and $w_*$ is the mean upward velocity.
For a more detailed discussion of how spherical particles are treated in {\virga}, see \citet{batalha2020,rooney2022,batalha2025virga}. To add treatment of fractal aggregates within this framework, we begin by defining aggregates as clusters of smaller particles (monomers) that are stuck together, which exhibit self-similarity over a finite range of length scales, as shown in Figure~\ref{fig:cartoon}. Fractal aggregates are well-described mathematically by:

\begin{equation} \label{eq:agg}
    N_\mathrm{mon} = k_0 \left( \frac{R_\mathrm{agg}}{r_\mathrm{mon}} \right)^{D_f},
\end{equation}

\noindent where $N_{\rm{mon}}$ is the number of monomers, $D_{\rm{f}}$ is the fractal dimension, $k_0$ is the prefactor discussed below, $R_\mathrm{agg}$ is the characteristic radius of the aggregate (here defined as the radius of gyration) and $r_\mathrm{mon}$ is the radius of the monomers. Highly linear aggregates are described by low fractal dimensions (i.e., $D_{\rm{f}} = 1$ would be a perfectly linear chain of monomers) while highly compact aggregates approach $D_{\rm{f}} = 3$, which would describe a perfect compact sphere. To compare spheres and aggregates in this paper, we follow the method in \citet{Lodge2024} and compare particles of equivalent mass:

\begin{align} 
    \begin{split}\label{Eq:M_agg}
        M_\mathrm{sphere} &= M_\mathrm{agg} 
    \end{split}
\end{align} 
\noindent and
\begin{align}
    \begin{split}\label{Eq:M_agg2}
        M_\mathrm{agg} = N_\mathrm{mon} m_\mathrm{mon} =
        \rho_{m}\frac{4}{3}\pi r^3 = N_\mathrm{mon}\rho_{m}\frac{4}{3}\pi r_\mathrm{mon}^3,
    \end{split}
\end{align} 

\noindent where $r$ is the radius of a sphere with equivalent mass to the aggregate, $m_{\rm{mon}}$ is the mass of a single monomer and $\rho_{m}$ is the bulk density of the material (the individual monomers are assumed to have no porosity). Here, this implies that, as in \citet{Lodge2024}:

\begin{equation} \label{eq:nmon}
        r^3 = N_\mathrm{mon} r_\mathrm{mon}^3 
\end{equation} 

The volume of the aggregate is calculated using the radius of gyration:

\begin{equation} \label{Eq:V_agg}
    V_\mathrm{agg}=\frac{4}{3}\pi R_\mathrm{agg}^3.
\end{equation}

\noindent Combining Eqs.~\ref{eq:agg}-\ref{Eq:V_agg}, we find that the aggregate density $\rho_{\rm{agg}}$ is given by:

\begin{equation}
    \rho_\mathrm{agg} = N_\mathrm{mon} \rho_{m} \left(\frac{r_\mathrm{mon}}{R_\mathrm{agg}}\right)^3 = k_{\rm 0}\rho_{\rm m}\left(\frac{r_\mathrm{mon}}{R_\mathrm{agg}}\right)^{(3-D_{\rm f})}.
\end{equation}

\begin{figure*}
\begin{center}
\includegraphics[width=0.8\textwidth]{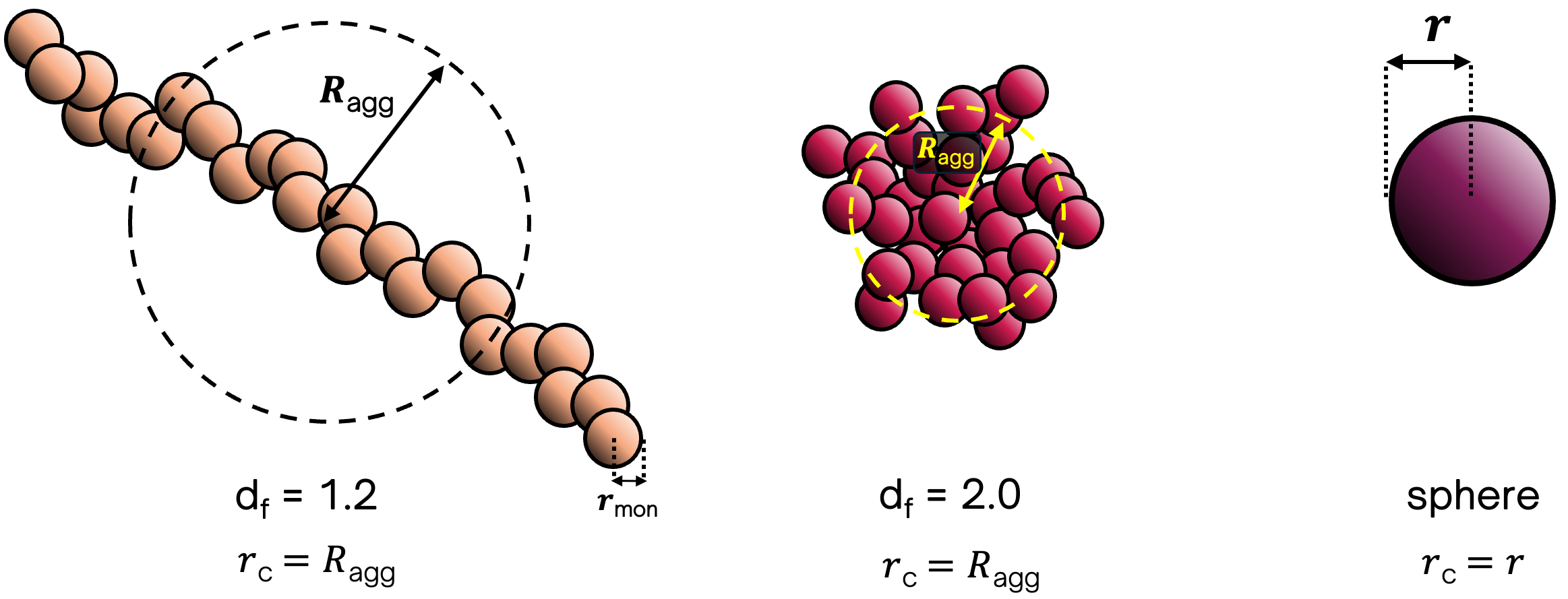}
\caption{The variety of different radii definitions that are used in this paper for spherical and non-spherical shapes. To illustrate the definitions, in the example above all three shapes above have identical mass, but the characteristic radius $r_{\rm{c}}$ (used for calculating the optical properties and the fall velocity) is different for each shape. The characteristic radius $r_{\rm{c}}$ for a sphere is $r$. The characteristic radius for an aggregate is $R_{\rm{agg}}$, which is taken to be equal to the radius of gyration $R_{\rm{gyr}}$ (and as shown above, this varies depending on the distribution of mass). $R_{\rm{agg}}$ is much larger for linear, lacy shapes compared to more compact shapes, even when the masses of the two aggregates are the same. The compact radius $r_{\rm{agg,m}}$ of the aggregates would be equal to $r$ because the masses of these three shapes are the same (i.e., if the aggregates were compressed into spheres, all three spheres would have the same radius $r$). The monomer radius $r_{\rm{mon}}$ is also shown. The radius $r_{\rm{w}}$ (see Eq.~\ref{eq:alpha}) represents the characteristic radius where $v_{\rm{fall}}=w_*$ and this would therefore be different for all three shapes. 
}
\label{fig:radii_definitions}
\end{center}
\end{figure*}

Particle movement through an atmosphere depends on the size and density of the particle, the density and viscosity of the atmosphere, and the relative flow speed between the two. The characteristic dynamical radius ($r_{\rm{c}}$) for a sphere is simply its geometric radius ($r$). We follow the suggestions in \citet{ohno2020} by using the radius of gyration $R_\mathrm{gyr}$ for aggregates \footnote{{Note that some studies define the characteristic radius as the radius of a sphere with the equivalent gyration radius, $\sqrt{5/3}R_{\rm agg}$. See e.g., \citet{Tazaki2021} for definitions of various radii used to characterize fractal aggregates.}}, which gives ideal scaling with aggregate size (see \citealt{sorensen2011mobility}), and also keeps consistent with the optical model used (see Section \ref{sec:optics}). All references to $R_\mathrm{agg}$ in this paper therefore refer to the radius of gyration of the aggregate ($R_\mathrm{agg}=R_\mathrm{gyr}$).  We also define $r_{\mathrm{agg,m}}$ as the radius of an aggregate if it were compressed into its equivalent compact mass spherical form (to allow us to more directly compare how much condensate material forms into each shape type). To clarify the various radii definitions used in this paper, see Figure~\ref{fig:radii_definitions}.

To determine the flow regime, we first determine the Knudsen number, Kn:

\begin{equation}
     \mathrm{Kn} = \frac{\lambda}{r_{\rm{c}}}
\end{equation} 

\noindent where $r_{\rm{c}}$ is the characteristic radius of the particle ($r$ for spheres, $R_\mathrm{agg}$ for aggregates) and $\lambda$ is the mean free path of the atmosphere, defined as

\begin{equation}
     \lambda = \frac{1}{\sqrt{2} n_a \pi d^2}
\end{equation} 

\noindent and where $n_a$ is the number density of molecules in the atmosphere and $d$ is the molecular diameter. (In a hydrogen-dominated atmosphere, $n_a$ is equivalent to $\sim 2\mu \sim 3.3\times10^{-24}$ g via the ideal gas law and $d$ is $2.827\times10^{-8}$ cm).

If Kn $\gg1$, particles are in the free molecular regime, where the mean free path of the gas is large compared to the particle size, and so drag is modelled as individual gas molecules colliding with the particle. In this regime, the terminal velocity of a spherical particle is given (e.g., {\eddysed}, \citealt{ohno2020}, \citealt{sanaz2024}) by:

\begin{eqnarray} \label{eq:epstein}
    v_\mathrm{fall, sphere}&=& \frac{\frac{2}{3}g \rho_{m}}{v_\mathrm{th}\rho_{a}}r
\end{eqnarray}

\noindent where $g$ is the acceleration due to gravity, $v_{\rm{th}}$ is the local speed of sound, and $r$ is the radius of the sphere. 

See \citet{sanaz2024} for the full derivation of Equation~\ref{eq:epstein}, and note that other works simplify the expression differently, e.g., \citet{woitkeandhelling2003}. For fractal aggregates in the free molecular regime, we can use Eq. 46 of \citet{sanaz2024}:

\begin{eqnarray}\label{eq:v_agg_sanaz2024}
    v_\mathrm{fall, agg} &=& v_\mathrm{fall, sphere} \left(\frac{R_\mathrm{agg}}{r_\mathrm{mon}}\right)^{(2D_f-6)/3}.
    \\ \nonumber
    &=&\frac{2g\rho_{\rm m}}{3v_{\rm th}\rho_{\rm a}}k_{\rm 0}^{-(2D_{\rm f}-6)/3D_{\rm f}}r_{\rm mon}N_{\rm mon}^{(D_{\rm f}-2)/D_{\rm f}},
\end{eqnarray}
where we have used Equations \eqref{eq:agg} and \eqref{Eq:M_agg2} for the second line to express fall speed as a function of $N_{\rm mon}$ and $r_{\rm mon}$.
        
If Kn $\ll1$, the particle is in the continuum regime (where the gas is treated as a continuous fluid). Between the extremes of Kn $\ll1$ and Kn $\gg1$, particles are in the transitional flow regime. Within our framework, we assume a value of Kn = 10 as the bifurcation of the free to continuum regimes. Here, we use Eq. 23 of \citet{Ohno2017} to calculate $v_\mathrm{fall}$ for any particle shape in the continuum regime:

\begin{equation} \label{Eq:v_fall}
    v_\mathrm{fall} = \frac{2\beta g r_c^2 \rho_{c}}{9\eta} \left[ 1 + \left( \frac{0.45g r_c^3 \rho_a \rho_c}{54\eta^2} \right)^{2/5} \right]^{-5/4}
\end{equation}

\noindent where $r_c$ is the characteristic radius of the particle ($r$ for spheres, $R_\mathrm{agg}$ for aggregates), $\rho_c$ is the characteristic density of the particle ($\rho_{m}$ for spheres, $\rho_\mathrm{agg}$ for aggregates), $\eta$ is the atmospheric viscosity, and the slip correction factor $\beta$ is given by (Eq. 8 of \citealt{ohno2018microphysical}):

\begin{figure*}
\begin{center}
\includegraphics[width=0.8\textwidth]{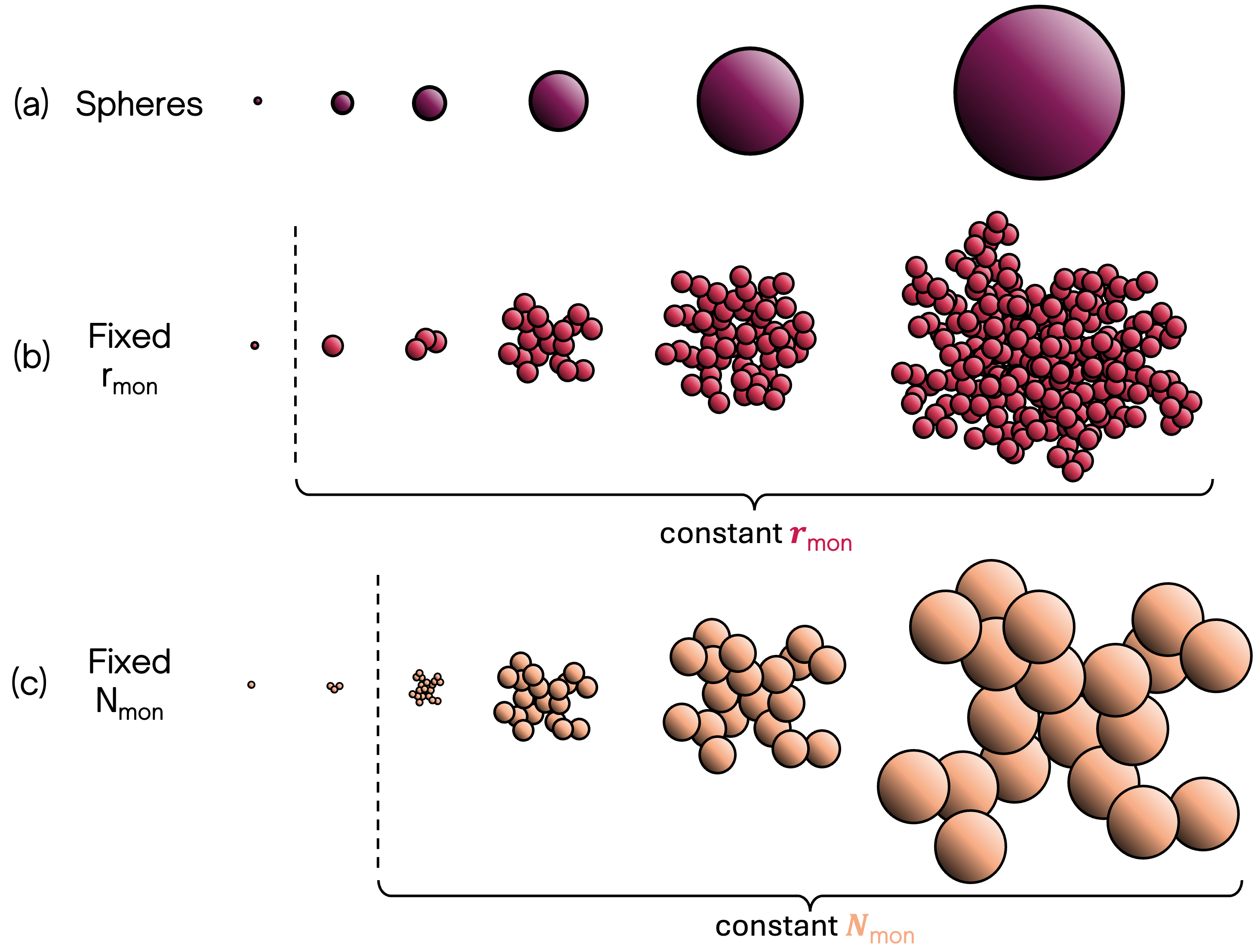}
\caption{Schematic of aggregate particle options as implemented into {\virga}. (a) Spheres grow in radius as larger particles are needed to balance higher convective velocities in Eq.~\ref{Eq:eddysed}. (b) If the user prescribes a fixed monomer radius $r_\mathrm{mon}$: as particles grow, only the number of monomers increases. However, to allow smaller particle solutions in lighter convective velocity situations, very small particles are treated as single monomers until they are large enough to have a radius equal to $r_{\rm{mon}}$ (this region is shown to the left of the dashed line). (c) If the user prescribes a fixed number of monomers $N_{\rm{mon}}$, the radius of each of the monomers increases as particles grow. Similarly to case (b), smaller particle solutions in lighter convective velocity situations are allowed by treating aggregates as collections of smaller numbers of monomers (with the smallest possible physical monomer radius, shown left of the dashed line) until the set $N_{\rm{mon}}$ is reached.}
\label{fig:cartoon}
\end{center}
\end{figure*}

\begin{equation}
    \beta = 1 + \mathrm{Kn} \left(1.257 + 0.4\exp \left(-\frac{1.1}{\mathrm{Kn}} \right) \right).
\end{equation}

Eq.~\ref{Eq:v_fall} is valid for spheres and aggregates, for all three of the Stokes, slip and turbulent regimes. It should be highlighted that \citet{ohno2020} assumes $D_{\rm{f}}$ = 2 throughout, and Eq.~\ref{Eq:v_fall} is not strictly designed to work for very low fractal dimensions (e.g., diffusion-limited cluster aggregates (DLCA) with $D_{\rm{f}} = 1.2$), so results for these shape types should be considered as an upper bound. In \citet{ohno2020} and \citet{sanaz2024}, the fractal prefactor was set at $k_0=1$ for all particles. Here, we use the linearly interpolated expression proposed by \citet{Tazaki2021}, which captures the behavior of aggregates formed from cluster-cluster aggregation to linear-chains, valid for 1 $\le D_{\rm{f}} \le$ 3 (to within 5 \% accuracy) for $N_{\rm{mon}}\geq100$ monomers:

\begin{equation} \label{eq:k_0}
    k_0 = 0.716(1 - D_{\rm{f}}) + \sqrt{3} 
\end{equation}

To retain accuracy at low $N_{\rm{mon}}$ (to better represent particle growth from single monomers), we create a linear fit below $N_{\rm{mon}}<100$ monomers such that $k_0$ smoothly transitions to $k_0=1$ at $N_{\rm{mon}}=1$ for any shape type:

\begin{equation}
    k_0 = \frac{1.448-0.716D_{\rm{f}}}{99}N_{\rm{mon}} + \left(1 - \frac{1.448-0.716D_{\rm{f}}}{99} \right).
\end{equation}

Where we use the term ``particle growth'', it is important to highlight that our parametrized model does not track particle growth through the atmosphere, or microphysical processes such as particle compression or ballistic destruction (see \citealt{ohno2020}). Instead, {\virga} sets the particle size by balancing convection and fall speed for a particular pressure layer. Then all particles in that layer are made into a particular shape, based on one of two user-prescribed models. We consider two simple methods of building fractals:

\begin{enumerate}
    \item Increase the number of monomers ($N_{\rm{mon}}$), but keep the monomer size ($r_\mathrm{mon}$) fixed.
    \item Fix the number of monomers ($N_{\rm{mon}}$), but increase the monomer size ($r_\mathrm{mon}$).
\end{enumerate}

In the code, users choose a value for \texttt{Df}, and also choose whether to fix either \texttt{r\_mon} or \texttt{N\_mon} -- these two growth options are shown in Figure \ref{fig:cartoon} (b) and (c) respectively. In reality, aggregates would probably grow in a manner somewhere between these two models, but being able to implement them separately provides valuable insight into the effects and consequences of each type of growth process, whilst maintaining the simplicity and computational speed of {\virga}. All particles in a particular layer are considered to have grown into the same shape type, but their size within each layer still varies as a log-normal distribution in the same way as the original model \citep{ackerman2001,batalha2025virga}. 

To close the system analytically, \citet{ackerman2001} prescribe a simple power-law dependence relating the fall speed of a spherical particle and its radius:

\begin{equation} \label{eq:alpha}
    v_\mathrm{fall,sphere} = w_* \left( \frac{r}{r_w} \right)^\alpha
\end{equation}

\noindent where $r_w$ is the radius of a particle that has a terminal velocity equal to the convective upwards velocity (in other words, where $v_\mathrm{fall,sphere}(r)=w_*$). The exponent $\alpha$ then acts as the link between $r_w$ and the geometric radius of particles in a lognormal distribution of width $\sigma$ (see \citealt{ackerman2001}), from which the optical properties of the distribution (see Section \ref{sec:optics}) can be determined:

\begin{equation} \label{eq:r_geo}
    r_g = r_w f_\mathrm{sed}^{1/\alpha} \exp\left( - \frac{\alpha+6}{2} \ln^2 \sigma \right).
\end{equation}

\begin{figure*}
\centering
{\includegraphics[width=0.9\textwidth]{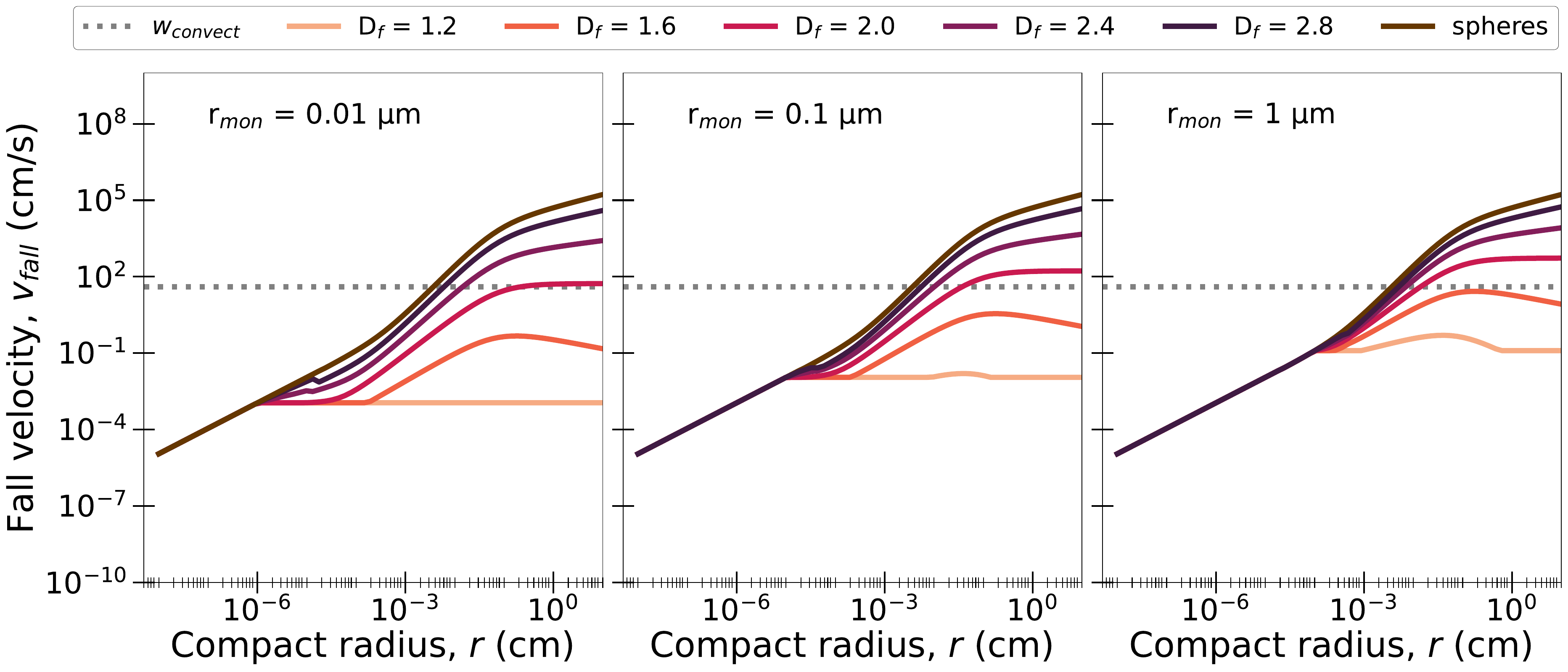}
\includegraphics[width=0.9\textwidth]{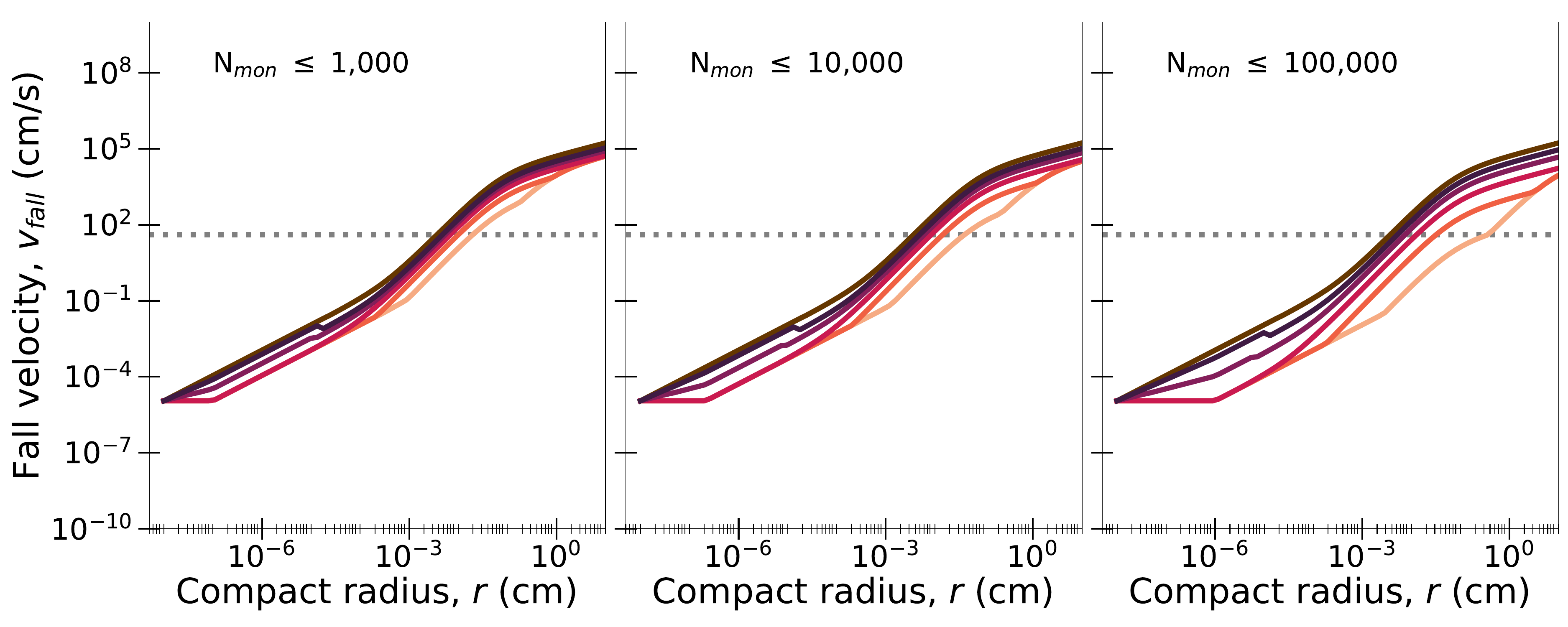}}
\caption{Fall speeds, $v_{\rm{fall}}$, of compact spheres with radius $r$ compared to aggregate particles (in terms of their equivalent compact mass radius $r_{\rm{agg,m}}$) of varying fractal numbers ($D_{\rm{f}}$) at single layer (pressure of 0.1 bar) of an example model atmosphere. At the radius where the fall speed is equal to the convective velocity ($w_{\rm{convect}}$; dashed grey line), a particle of that radius would be stable in that atmospheric layer under the {\virga} parametrization. Lower $D_{\rm{f}}$ (i.e., more linear) aggregates grow to larger radii before reaching high fall speeds. Top row shows our implementation where we set the size of the monomer radius $r_{\rm{mon}}$; bottom row shows when instead we set the number of monomers, $N_{\rm{mon}}$. For low fractal dimensions, the aggregate fall speed is tied to the monomer radius fall speed (see text). 
}
\label{fig:vfall}
\end{figure*}

The above works for spherical particles, but fractal aggregates can have much more complex dependencies for $v_{\mathrm{fall}}$ as a function of $r$ than the power-law dependence in Eq.~\ref{eq:alpha}. For very fluffy aggregates with $D_{\rm{f}}<2$ and fixed monomer sizes, $v_{\mathrm{fall}}$ as prescribed by Eq.~\ref{Eq:v_fall} can decrease as the overall particle size grows. An aggregate's cross section, however, cannot exceed the sum of its constituent monomer cross sections \citep{Tazaki2021}, which is implicitly assumed in Eq.~\ref{Eq:v_fall} for $D_{\rm{f}}>2$. When $D_{\rm{f}}<2$, this physical constraint is violated. Therefore, without adjustment, our formulation can produce much smaller fall velocities than is realistic. To counteract this behavior,  we impose the condition that $v_{\rm{fall}}$ of any aggregate $>$ $v_{\rm{fall}}$ of its constituent monomers, which is shown in Figure \ref{fig:vfall}.

Moreover, there are cases with no aggregate solutions to Eq.~\ref{Eq:eddysed} with Eq.~\ref{eq:alpha} -- for example, a user may prescribe a low $K_{zz}$ (the equivalent of very light upwards convective conditions), and a fixed $r_\mathrm{mon}=1~\mu$m, but a single monomer particle may have a fall speed that is greater than the convective upwards velocity. Therefore, in these cases, spherical particles smaller than the prescribed $r_\mathrm{mon}$ are allowed. Similarly, if a user prescribes $N_{\rm{mon}}=1000$, the smallest aggregates that could be formed are 1000 of the smallest physical particles in the model, which again may be too ``heavy'' to balance upwards convection, so smaller numbers of monomers are allowed here too. Both of these model growth options, including the exceptions discussed above, are shown in Figure~\ref{fig:cartoon}.

Due to the above complexities in calculating $v_{\rm{fall}}$ for a wide range of aggregate shapes and sizes, we cannot follow the simple prescription in Eq.~\ref{eq:alpha} under our aggregate implementation. Instead $\alpha$ is determined using spherical particles (using Eq.~\ref{eq:alpha}), and we assume the same relationship between $r_g$ and $r_w$ holds for aggregates.

Figure \ref{fig:vfall} shows how the fall speed of a particle varies as the radius increases, for both versions of our implementation: either 1) setting the particle size (fixed $r_{\rm{mon}}$) or 2) setting the number of monomers (fixed $N_{\rm{mon}}$). As expected, smaller monomers fall more slowly than larger monomers. As a consequence of the relationship between the number of monomers and the radii of monomers -- i.e., Eq.~\ref{Eq:M_agg2} and \ref{eq:nmon} -- larger numbers of monomers under fixed $N_{\rm{mon}}$ are made up of smaller monomer radii, and also fall more slowly compared to small fixed $N_{\rm{mon}}$, which are made up of larger monomer radii. This relationship can be seen by comparing the last column of Figure~\ref{fig:cartoon}, where the same outer aggregate in the fixed $N_{\rm{mon}}$ is made of large monomers, while the fixed $r_{\rm{mon}}$ has many more monomers that are smaller.  These effects are heightened as the fractal dimension becomes lower and lower, in which particles are getting less compact and more lacy.

For some cases where $D_{\rm{f}}<2$, the $v_{\rm{fall}}$ curve flattens as the calculated fall velocity approaches that of a monomer, where we impose our strict lower limit, seen in Figure~\ref{fig:vfall}. In this instance, there can exist multiple solutions (multiple particles of size $r_w$ that could balance $w_*$), in which case we use the smallest solution. Additionally, cases can occur where no solution exists if the user-prescribed convective velocity $w_*$ is too high and $r_\mathrm{mon}$ is fixed; the particle may be ``too fluffy'' to stay in the pressure layer. In such cases, the flattening or negative inversion of the $v_\mathrm{fall}$ equation where $D_{\rm{f}}<2$ (see top panel of Figure~\ref{fig:vfall}) means that $v_\mathrm{fall}$ may never be high enough to balance $w_*$. 
The occurrence of such a situation can also be inferred from Equation \eqref{eq:v_agg_sanaz2024}, which indicates that the terminal velocity cannot increase with increasing $N_{\rm mon}$ for aggregates with $D_{\rm f} \le 2$ in the kinetic regime (${\rm Kn}>10$).

In a microphysical model, such fluffy particles would rise through the atmosphere, but we strive to maintain the ethos and simplicity of the original {\virga} model here and withhold from implementing bulk movement of condensate. Instead, {\virga} gives an error and advises the user to decrease the eddy diffusion coefficient $K_{zz}$ (and therefore decrease $w_*$). This effect only happens for cases of high $K_{zz}$ and where $r_\mathrm{mon}$ is fixed -- if $N_{\rm{mon}}$ is fixed instead, the monomers can grow arbitrarily large and eventually $v_\mathrm{fall}$ will balance $w_*$ at some point (see bottom panel, Figure~\ref{fig:vfall}), so a solution will always be found.

At the limits of this model, if the user prescribes a $K_{zz}$ (and therefore an upwards convective velocity $w_*$) that is very low, it is possible to have atmospheric conditions where particle sizes would need to be smaller than atoms to balance convection (this is not unique to fractal aggregates, and can happen for spherical particles as well). To guard against this, fundamental physical limits have been put in place, and users are advised to increase $K_{zz}$ if solutions cannot be found and the code throws an error. These fundamental physical limits apply to spheres as well -- we do not allow particles to form that are smaller than atoms (i.e., particles must be at least 1 \AA\ in radius). 
Users are always advised to check that their model outputs are reasonable, and we provide several tutorial-style notebooks in the online documentation\footnote{\href{https://github.com/natashabatalha/virga/blob/0377aa8803ec757f592cf4667c291bea7138f34e/docs/notebooks/10_Fractal_Aggregate_Aerosols_in_a_Hot_Jupiter.ipynb}{Virga 2.0 Tutorials; \citetalias{virgaV2batalha}}} to demonstrate various checks.




\section{Optics of Fractal Aggregate Particles}
\label{sec:optics}

With the dynamics of falling fractal aggregates now self-consistently accounted for in {\virga}, we next describe our method for computing the optical properties of these aggregate particles. In the spherical version, {\virga} uses classic Mie Theory \citep[i.e.,][]{bohrenhuffman1983} to compute the asymmetry parameter, single scattering albedo, and optical depth as a function of wavelength and particle size. Many open-source codes exist to compute Mie coefficients for spheres, but extensive vetting resulted in {\virga} Version-0 using the \texttt{PyMieScatt} package \citep{piemiescatt} and {\virga} Version 1.0 \citep{batalha2025virga} using  \texttt{miepython} \citep{miepython}. 

To expand into the optical behavior of fractal aggregates, we implement Modified Mean Field Theory as developed by \citet{Tazaki2018} and \citet{Tazaki2021}, which was also used in the exoplanet aggregate cloud model of \citet{ohno2020}. Modified Mean Field Theory (MMF) is able to calculate the required optical properties of fractals that are described by the same parameters as in Equation \ref{eq:agg}, assuming a conglomerate particle made of self-similar (i.e., all the same size and the same material) monomers. MMF accounts for singly and multiply scattered inter-monomer waves within an aggregate particle while also considering the overall aggregate's shape (i.e., its fractal dimension, $D_{\rm{f}}$). Thus, MMF captures the extinction, scattering, and absorption across a wide wavelength range more accurately than Mie theory or effective medium theory (EMT) coupled to Mie theory.

There are some very rigorous and accurate approaches to computing the optical properties of non-spherical particle shapes, such as the Discrete Dipole Method (DDA) \citep{purcell1973scattering, draine1994discrete}. The DDA is very flexible, and capable of calculating properties of any particle shape by breaking it into a system of smaller dipoles, and then computing the overall scattering and absorption. It can even calculate optical properties for heterogeneous mixtures of materials using their exact relative embedded locations within a larger aggregate (see \citealt{adachi2010shapes}), making it a very flexible theory, but much more computationally intensive. Unfortunately, its slow speed makes it frequently prohibitive for large parameter space sweeps (i.e., many particle radii, wavelengths, and shape types), though recent work has explored parameter spaces where DDA can be empirically approximated extremely quickly \citep{lodge2024manta}. However, for general parameter spaces, DDA calculations are still expected to take significantly longer than other approaches. A full comparison of Mie theory, MMF and DDA for fractal aggregates can be found in the \texttt{CORAL} code publication \citep{Lodge2024}. Additional discussion is also found in \citet{sanaz2024} and \citet{kiefer2024}. 

We benchmark our results in this paper against DDA, but to balance both speed and accuracy for a wide range of parameter space, we choose to use MMF in the general release of {\virga}. We use the open source package \texttt{Optool} \citep{dominik2021}, which implements the method using the original theory developed by \citet{Tazaki2018} and \citet{Tazaki2021}. This code is written in Fortran 90 but includes a Python wrapper. In our use of \texttt{Optool}, the parameter ``\textit{iqcor}'' represents the cutoff function assumed in the two-point correlation function \citep{Tazaki2018}. \citet{Tazaki2021} suggests this value should be set to 3 (imposing a fractal dimension cutoff) in the fractal domain where D$_{\rm{f}}$ $\leq$ 2, and to 1 (imposing a Gaussian cutoff) for more compact aggregates with larger fractal number. However, the \texttt{Optool} documentation suggests always keeping this parameter set to 1 for numerical stability. We examined both implementations and found $\sim$ 1-2\% differences in the resulting scattering and extinction efficiencies. In addition, setting \textit{iqcor} = 3 doubles the computation times. Therefore, we ultimately leave \textit{iqcor} set to 1, with a Gaussian cutoff, for all fractal numbers in our grid, as recommended.

\begin{figure}[ht!]
{\includegraphics[width=0.49\textwidth]
{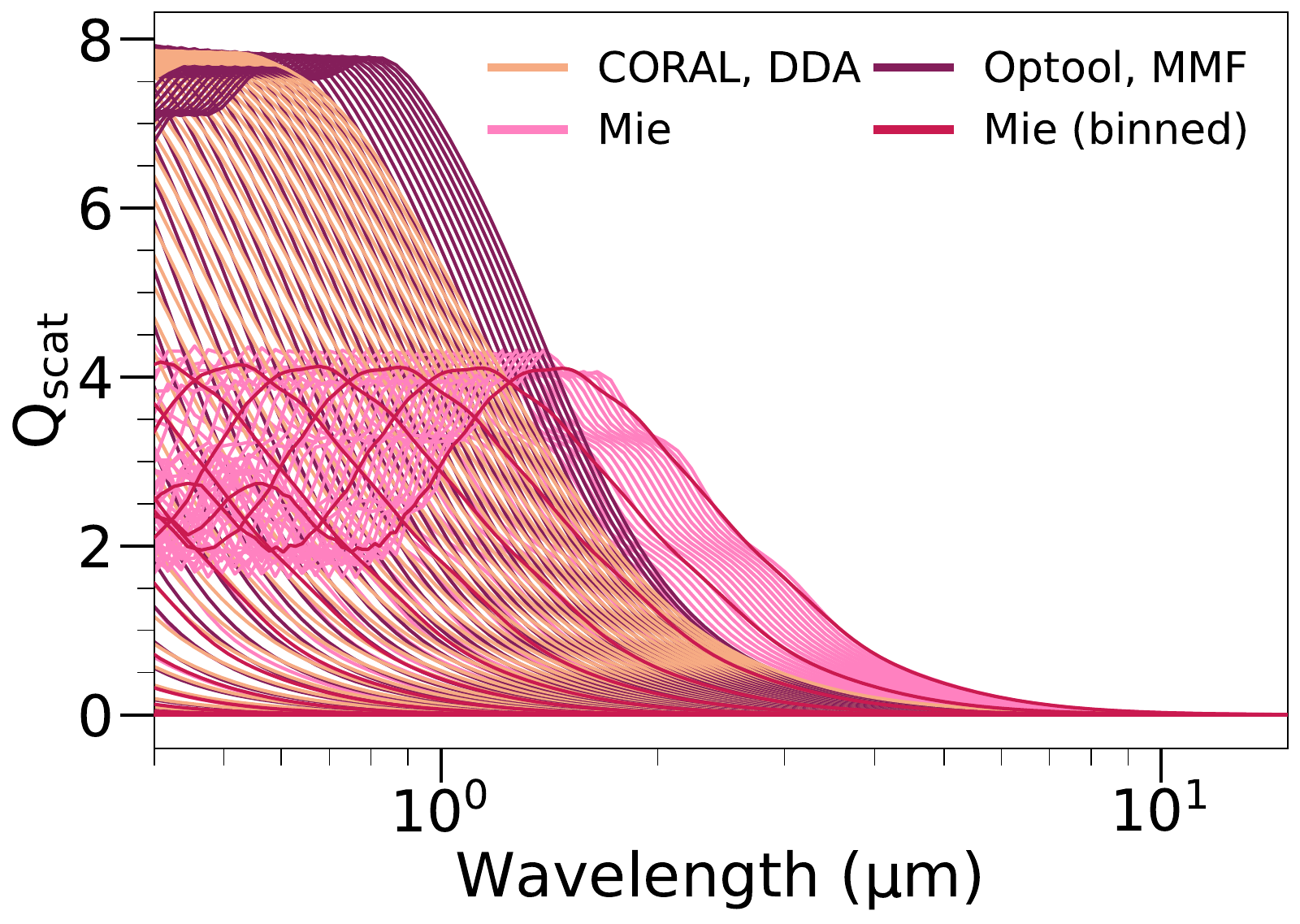}
\caption{Comparison of $Q_{\rm{scat}}$ shown for a fractal dimension $D_{\rm{f}} = 1.8$ over a grid of particle sizes using different optics tools, i.e.,  modified mean field theory (MMF) via \texttt{Optool} \citep{dominik2021}; the discrete dipole approximation (DDA) via \texttt{CORAL} \citep{Lodge2024}, as well as Mie theory from both. At a fraction of the computational cost, MMF theory is much closer to the ``reality'' of DDA compared to Mie theory for fractal aggregates.}
\label{fig:coral-comparison}}
\end{figure}

In Figure \ref{fig:coral-comparison}, we show a comparison between the outputs for the scattering efficiency, $Q_{\rm{scat}}$, for a grid of lacy ($D_{\rm{f}}$ = 1.8) KCl aggregates, as computed by both \texttt{CORAL} and \texttt{Optool} with Mie, MMF, and DDA (using the CLDR prescription of DDA, and the same comparison method detailed in \citealt{Lodge2024}). We compared the optical properties obtained for a variety of shapes and particle sizes, and the results from this figure are illustrative of results for all tests performed. The \texttt{Optool} run for MMF for this grid (with equivalent-volume spherical radii $r$ from 0.01 to 1 $\rm{\mu}$m, and 100 log-spaced wavelength values from 0.4 to 28 $\rm{\mu}$m) of KCl took approximately 10 minutes on a single laptop core, while the \texttt{CORAL} DDA run took 25 hours on a high performance computing cluster, though additional optimizations are likely possible. 

Using \texttt{Optool}, we generate a new MMF-based database for the scattering, absorption, extinction, and asymmetry parameters for aggregate particles. Since fractal aggregates are in part defined by their aggregate radius, $R_{\rm{agg}}$, we use a subset of the existing radii grid already within spherical {\virga} as our range of aggregate particle radii in terms of the equivalent compact mass radius $r_{\rm{agg,m}}$, which currently ranges from $r$~=~1$\times$10$^{-8}$~cm to 0.5 cm for spheres in 60 steps. Our aggregate grid covers  1$\times$10$^{-7}$~cm to 0.01~cm in 40 steps, as we find in practice this covers all typical solutions to the stable size of $R_{\rm{agg}}$ discussed in Section \ref{sec:dynamics}. Users are advised to check that the radii solutions found by {\virga} do have a counterpart in the MMF grid; a warning will be raised when running the code if not.

\begin{figure*}
\begin{center}
\includegraphics[width=0.99\textwidth]{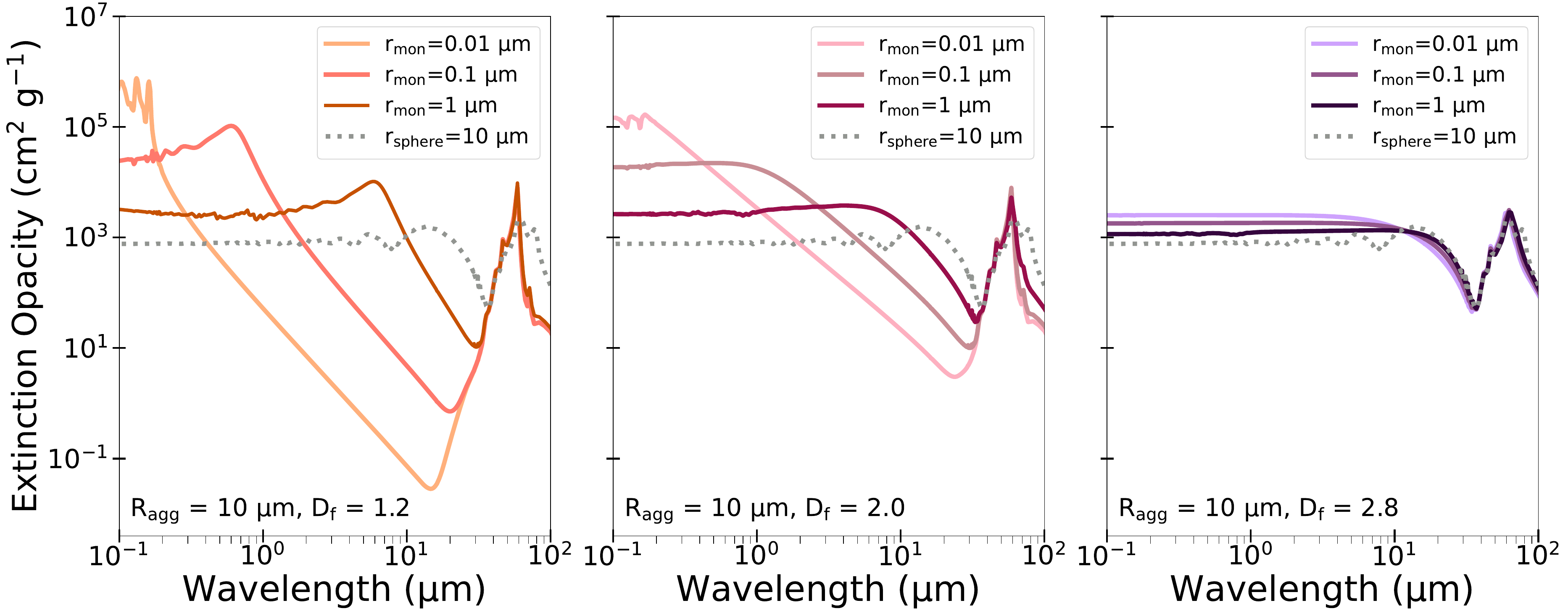}
\caption{The extinction opacity of aggregate particles of KCl as a function of wavelength for different monomer sizes $r_{\rm{mon}}$ with an aggregate radius of $R_{\rm{agg}}$ = 10 $\mu$m, computed by MMF theory with \texttt{Optool}. The dotted grey line shows extinction for a spherical particle with a radius of 10 $\mu$m. From left to right, we show fractal dimensions $D_{\rm{f}}$ of 1.2, 2.0, and 2.8. Our middle panel with $D_{\rm{f}}$ = 2 reproduces Figure 4 of \citet{ohno2020}.}
\label{fig:opacity}
\end{center}
\end{figure*}

As shown in the Mie extinction curves in Figure \ref{fig:coral-comparison}, Mie theory results in resonance fringes for the scattering of a single particle \citep{Hansen1974}. Thus, {\virga} includes a sub-bin radii smoothing step \citep{batalha2025virga} from the outputs of 
\texttt{miepython}, which we also apply to \texttt{Optool} outputs as shown in Figure \ref{fig:coral-comparison} to average out over these fringe effects. Since all implementations of {\virga} always compute lognormal size distributions of particles, this method avoids non-physical scattering or optical depth resonances due to single particles. {\virga} then uses the solutions found from Eq.~\ref{Eq:eddysed} and Eq.~\ref{eq:alpha} as the values to use for $R_{\rm{agg}}$, along with the width of the lognormal size distribution as set by $\sigma$ from Eq.~\ref{eq:alpha} and \ref{eq:r_geo}. 

\texttt{Optool} outputs $\kappa_{\rm{ext,sca,abs}}$ values, or mass opacities, rather than scattering, absorption, or extinction efficiencies, $Q_{\rm{ext,sca,abs}}$, as required by {\virga}. Therefore, we must compute efficiencies from \texttt{Optool} by the conversion, as in for example, \citet{Min2015}:

\begin{equation}
   \kappa_{\rm{ext,sca,abs}} = \frac{C_{\rm{ext,sca,abs}}}{M}
\end{equation}

and 

\begin{equation}
   Q_{\rm{ext,sca,abs}} = \frac{C_{\rm{ext,sca,abs}}}{\pi  r^2}
\end{equation}

\noindent where \textit{C$_{\rm{ext,sca,abs}}$} are the effective cross sections and \textit{M} is the total mass of the particle. By substituting the monomer density, \textit{$\rho$}, and volume of the spherical particle, we arrive at 

\begin{equation}
   \kappa_{\rm{ext,sca,abs}} = \frac{3 Q_{\rm{ext,sca,abs}}}{4 \rho r}
\end{equation}

\noindent in order to calculate our efficiency parameters. Note that the $r$ here refers to a compact sphere; where we compute aggregates, we use the term $r_{\rm{agg},m}$ to refer the aggregate in terms of the mass equivalent radius of a compact sphere. As discussed in Section \ref{sec:dynamics}, our code includes two options for generating fractal aggregates by either setting the number of monomers, $N_{\rm{mon}}$ or the size of monomers, $r_{\rm{mon}}$. For fixed $r_{\rm{mon}}$, our MMF grids include monomer radii $r_{\rm{mon}}$ = 0.01, 0.1, and 1 $\rm{\mu}$m. For fixed $N_{\rm{mon}}$, our grids include $N_{\rm{mon}}$ = 100, 1000, and 10000. For both sets of grids, we include fractal dimension $D_{\rm{f}}$ = 1.2, 1.6, 2.0, 2.4, and 2.8 for each set of condensate species included in the existing {\virga} distribution.

These grids of MMF-based optical properties can be found on Zenodo\footnote{\href{https://zenodo.org/records/16581692}{MMF Virga Opacities}; \citet{lodge_2025_16581692}} in the standard {\virga} format which can be directly coupled to the radiative transfer suite of {\picaso} for any user. Note that for the refractive index data required for our optical property computation, we use the KCl data from \citet{Palik1985} to better compare with the results presented in \citet{ohno2020} while the standard {\virga} version uses the data of \citet{querry1987optical}.

In generating our MMF grid, since an aggregate must have an aggregate radius larger than any individual monomer radius, we enforce the condition that if $R_{\rm{agg}}$ $\leq$ $r_{\rm{mon}}$, the Mie calculation reverts to \texttt{Optool}'s standard Mie theory for perfect spheres computation. The scattering and absorption from the smallest particle sizes even in our fractal aggregate treatment therefore likely result from perfect spheres in any given size distribution generated from {\virga}'s {\eddysed} function. This is consistent with our treatment of the aggregate particle fall speed discussed in Section \ref{sec:dynamics}.

In Figure~\ref{fig:opacity}, we show the extinction opacity, $\kappa_{\rm{ext}}$, for three of our fractal dimensions for the fixed $r_{\rm{mon}}$ version of our grid, for a single aggregate particle with an aggregate radius $R_{\rm{agg}}$ = 10 $\mu$m. The middle panel shows the same aggregate parameters as in \citet{ohno2020}. This figure demonstrates that as the difference between the monomer size and the aggregate particle size decreases, the particle extinction approaches that of a sphere. Larger monomer radii result in particles that have wavelength-independent extinction further into the infrared because the monomer size is large compared to the wavelength of light, while smaller monomers produce extinction that follows the Rayleigh limit to longer wavelengths. The intermediate regime where the wavelength of light lies between the size parameters of the monomer radius and the aggregate radius result in a non-Rayleigh slope produced by interference between individual monomers within the aggregate \citep{ohno2020}. Additionally, differences in particle extinction between monomer sizes are much more pronounced for lacy aggregates with $D_{\rm{f}}$ = 1.8 compared to more compact aggregates with $D_{\rm{f}}$ = 2.8 that approach nearly spherical shapes.

\newpage
\section{Case Study with Fractal Aggregate Clouds} \label{sec:results}

We now demonstrate the functionality of our new aggregate implementation in {\virga} using planetary parameters similar the well-studied sub-Neptune GJ~1214~b, using \ce{KCl} both with fluffy fractal particles and the standard spherical cloud particles. In this section, we compare to the results of previous studies by \citet{Adams2019} and \citet{ohno2020} who both investigated this planet, with fractal aggregate hazes and KCl clouds respectively. 

Both of these previous studies used microphysical cloud models, explicitly treating the nucleation, growth, and potential break-up or compression of particles via the \texttt{CARMA} (Community Aerosol and Radiation Model for Atmospheres; \citealt{GaoMarleyAckerman2018}) cloud code \citep{Adams2019} or a modified version of the model of \citet{Ohno2017}. As such, they were able to explore only a more limited parameter space, with only one fractal dimension for large aggregates. \citealt{ohno2020} used $D_{\rm{f}} = 2.0$ for all particles, with three values for the monomer radius. \citet{Adams2019} computed $D_{\rm{f}}$ as a function of the number of monomers increasing from $D_{\rm{f}} =1.5$ for two monomers to $D_{\rm{f}} =2.4$ for particles with $>$2000 monomers. Here, we explore a full range of potential fractal dimensions from $D_{\rm{f}} = 1.2$ to 2.8 which can be combined with any particle size. \citet{Adams2019} modelled fractal aggregate hazes rather than clouds, which form via photochemistry near the top of the atmosphere rather than via condensation at deeper pressures. Thus, the results of their study are less applicable as a point of comparison to our KCl clouds and indeed the current methodology of {\virga} generally, though we discuss the spectral consequences of our work compared to theirs in Section \ref{sec:spectra}.

GJ~1214~b, as the prototypical aerosol-ladden sub-Neptune, has been the subject of intense observational study, with transmission spectroscopy data from the optical to mid-infrared via ground-based, Hubble, and JWST observations \citep[e.g.,][]{bean2010,desert2011,kreidberg2014,kempton2023,schlawin2024}. Theoretical studies attempting to explain the muted spectrum of this planet are numerous, covering everything from photochemical hazes of varying compositions \citep[e.g.,][]{millerriccikempton,kawashima2018,lavvas2019,gao2023,ohno2024} to salt and sulfide clouds \citep[e.g.,][]{charnay2015cloud,gaobenneke2018,christie2022}, to combinations of both clouds and haze \citep{morley2013,morley2015,ohno2025,malsky2025}. 

The goal of this work is to enable comparisons with previous fractal aggregate modeling studies rather than explain the observational data of this world. Thus, we base our GJ~1214~b model atmosphere on the parameters used in \citet{ohno2020} rather than the most up-to-date constraints on the planet from more recent observational and interpretative work \citep[i.e.,][]{kempton2023,schlawin2024,ohno2025,malsky2025}. We do not provide comparisons to the existing data in our simulated transmission spectra (Section \ref{sec:spectra}). Our effort merely highlights the precision required to differentiate our models and that of the previous aggregate studies, rather than offering a new interpretation of the atmosphere of GJ~1214~b (see also, for example, \citealt{he2024}).

\subsection{Atmospheric Models with {\picaso}}
\label{sec:picaso}

First, we use the radiative-convective thermochemical equilibrium code {\picaso} 3.0 \citep{batalha2019,Mukherjee2023} to produce a model atmosphere in which to generate fractal aggregate clouds. We set the mass and radius of the planet to 0.244~$R_{\rm{Jup}}$ (2.73~$R_{\rm{\oplus}}$) and 0.0265~$M_{\rm{Jup}}$ (8.42~$M_{\rm{\oplus}}$), respectively. We use a stellar radius of 0.22 $R_{\rm{\odot}}$ with T$_{\rm{eff}}$~=~3500~K and log($g$)~=~5.0~cm~s$^{-2}$, with the planet at a semi-major axis of 0.14 au. 

\begin{figure*}
{\includegraphics[width=0.99\textwidth]{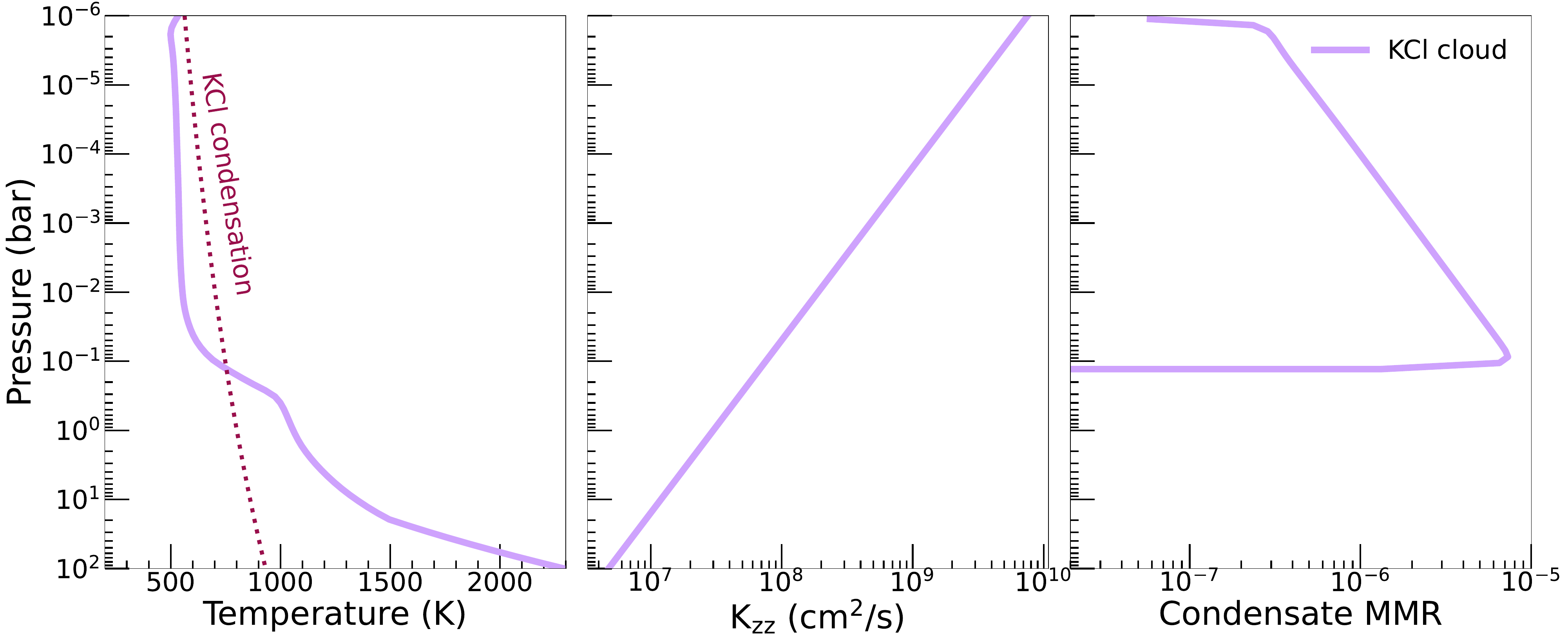}}
\caption{Left: Our GJ~1214~b-like model atmosphere pressure-temperature profile (solid, lavender) computed with \texttt{PICASO} for a 100$\times$ solar metallicity composition and the condensation curve of KCl (dashed, maroon) computed with {\virga}. Center: The eddy diffusion coefficient, $K_{\rm{zz}}$, evolution of our model atmosphere with pressure, following \citet{Charnay2015}. Right: the mean mass mixing ratio (MMR) of condensed KCl particles as produced by {\virga} with the model parameters, which is the same for all runs described in our case study, regardless of whether computed for spherical or aggregate particles.}
\label{fig:kzz}
\end{figure*}

We generate the temperature-pressure profile and chemistry with metallicity at 100$\times$ solar and the C/O ratio at its solar value, 0.458, following \citet{Lodders2009} using the correlated-k coefficients of \citet{lupu2021}. The \citet{lupu2021} pre-weighted opacities include 
\ce{C2H2}, \ce{C2H4}, \ce{C2H6}, \ce{CH4}, CO, \ce{CO2}, CrH, Fe, FeH, \ce{H2}, \ce{H3+}, \ce{H2O}, \ce{H2S}, HCN, LiCl, LiF, LiH, MgH, \ce{N2}, \ce{NH3}, OCS, \ce{PH3}, SiO, TiO, VO, Li, Na, K, Rb, and Cs. The intrinsic temperature is set to 200 K with heat redistribution factor at 0.6, representing near full day-night heat redistribution. We initialize the climate run using the \citet{Guillot2010} parametrization with 90 levels, an equilibrium temperature of 580 K, and upper and lower pressure bounds of 100 to 10$^{-6}$ bar. We set the number of convective zones at 1 and the initial guess of the uppermost layer of the convective zone at 82. We iterate on these initial guesses until the model reaches convergence, with the final temperature-pressure profile shown in the left panel of Figure \ref{fig:kzz}. These parameters were chosen to best match those of \citet{ohno2020}'s 100$\times$ solar metallicity runs.

Following \citet{ohno2020}, we also use the \citet{Charnay2015} parametrization for the eddy diffusion coefficient, $K_{\rm{zz}}$, for a 100$\times$ solar metallicity atmosphere on GJ~1214~b:

\begin{equation}
    K_{zz} = 3\times10^{7} (P)^{-2/5}
\end{equation}

\noindent where $P$ is the pressure in bars and $K_{\rm{zz}}$ is in cm$^2$ s$^{-1}$, shown in the center of Figure \ref{fig:kzz} for reference.


\subsection{Aggregate Cloud Properties in {\virga}}
\label{sec:cloud_properties}

With the atmospheric profile described above, we then run our new fractal aggregate version of {\virga} for KCl clouds. KCl condensation occurs as in \citet{morley2012}, which is shown in Figure \ref{fig:kzz}. Note that we do not adjust for the atmospheric metallicity in calculating its saturation vapor pressure profile due to the non-linear scaling of this dependence \citep{batalha2025virga}. The bulk density, $\rho_{\rm{m}}$, for KCl is 1.99 g cm$^{-3}$. We compute cloud profiles for each of the cases already discussed: 5 values for the fractal number $D_{\rm{f}}$: 1.2, 1.6, 2.0, 2.4, and 2.8; three values for fixed monomer radius $r_{\rm{mon}}$: 0.01, 0.1, and 1 $\mu$m; and three values for the fixed number of monomers $N_{\rm{mon}}$: 100, 1000, and 10000. All of our models were run with a lognormal width of $\sigma = 2$ and a sedimentation efficiency $f_{\rm{sed}}$ = 0.3, which is appropriate for a low-mass, warm sub-Neptune like GJ~1214~b \citep{morley2013,Brande2024}. We will explore the influence of both varying $K_{\rm{zz}}$ and $f_{\rm{sed}}$ on fractal aggregate clouds in future work (Lodge and Moran et al., in prep). 

\begin{figure*}
{\includegraphics[width=0.99\textwidth]{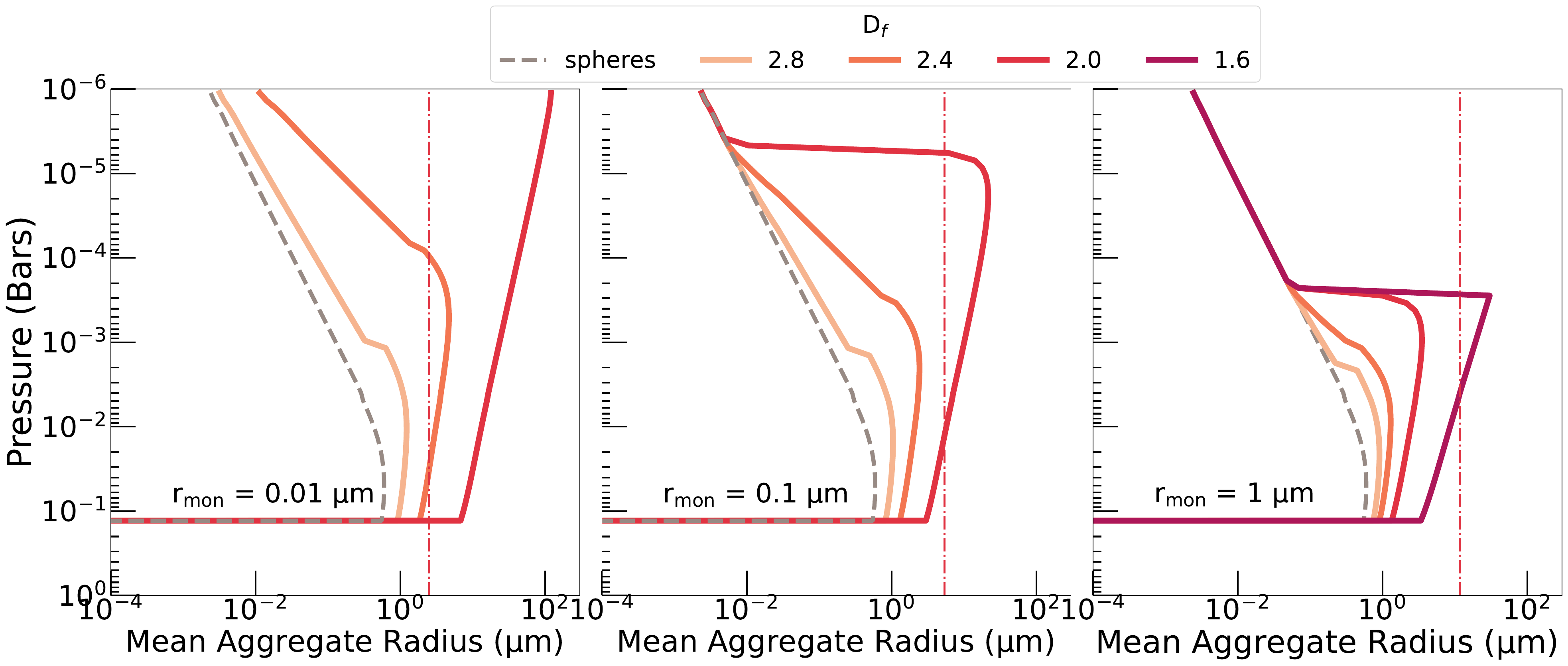}}
{\includegraphics[width=0.99\textwidth]{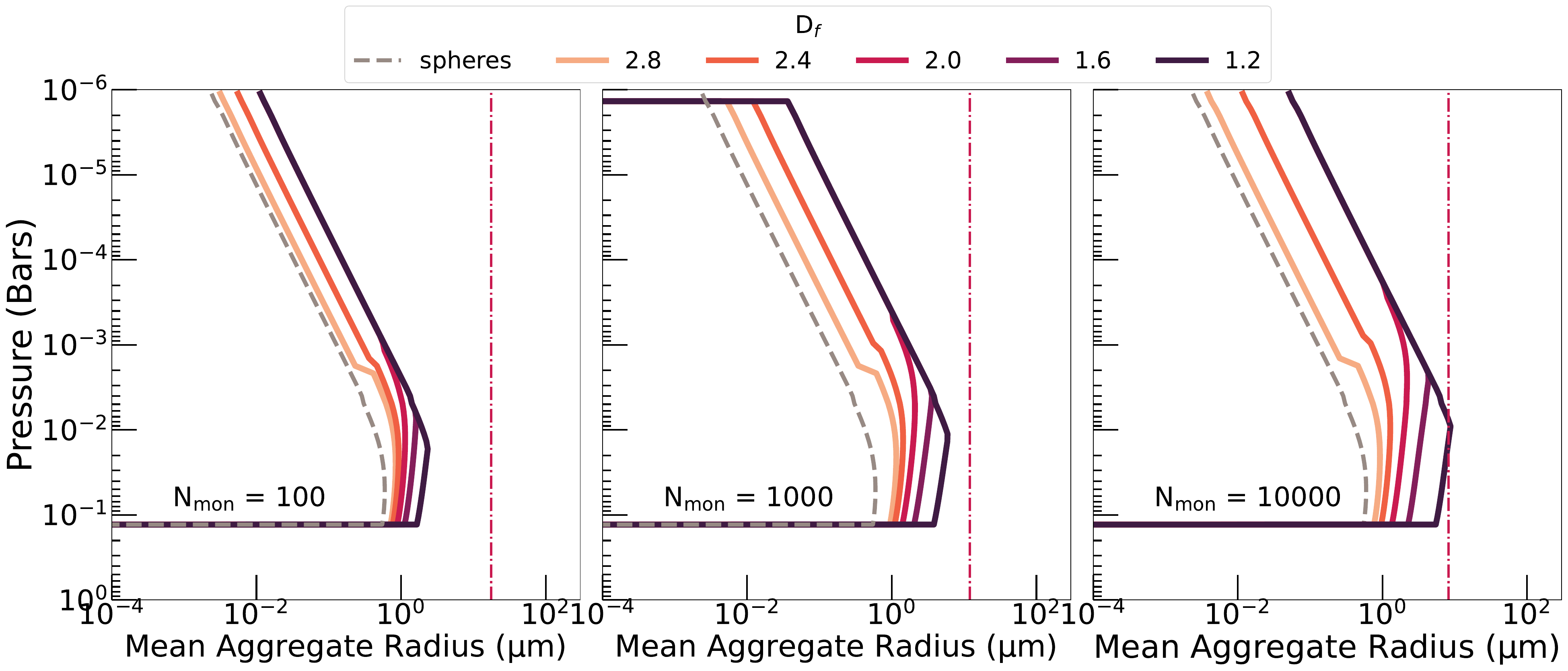}}
\caption{Top: mean particle size (in terms of equivalent mass compact spheres, $r_{\rm{agg},m}$) for fixed monomer sizes of $r_{\rm{mon}}$ = 0.01, 0.1, and 1 ${\rm{\mu}}$m as a function of pressure for our model GJ~1214~b atmosphere. Bottom: the same as top, but for fixed $N_{\rm{mon}}$ = 100, 1,000, and 10,000 monomers. The thin dash-dot red lines in both rows denotes the $D_{\rm{f}}=2.0$ compression radius taken from Equation 18 of \citet{ohno2020}, beyond which such particles are unlikely to persist (see text for more details).
}
\label{fig:particlesizes}
\end{figure*}

The mass of condensate formed in this atmosphere as function of pressure (or equivalently, altitude) is shown in the right panel of Figure \ref{fig:kzz}. Since our explicit assumption across all cloud particles -- spherical or aggregate -- is that condensate mass is conserved (Equation \ref{Eq:M_agg}), this condensate mass profile is the same across every single model run. This is a sharp contrast to the results of \citet{ohno2020}, wherein the cloud mass changes with each monomer size. The condensate at the cloud base at 0.1 bar in \citet{ohno2020} is approximately 10$^{-4}$ and decreases to 10$^{-8}$ at $\sim$10$^{-5}$ bar in the spherical case. Their aggregates with small monomers of $r_{\rm{mon}}$ = 0.1 and 0.01 $\mu$m increase the cloud mass to 10$^{-5}$ and 10$^{-4}$ mixing ratios, respectively, at the top of the model atmosphere. Our cloud for all particle shapes has a mass mixing ratio of only 10$^{-5}$ at the cloud base at 0.1 bar and 4$\times$10$^{-7}$ at the cloud top near the top of the atmosphere. At the cloud base, these differences stem from the fact that we do not change the saturation vapor pressure for KCl based on the metallicity of the atmosphere while \citet{ohno2020} do include this adjustment. Such saturation vapor trends are non-linear (see \citealt{batalha2025virga}, for further discussion) and beyond the scope of this model version.  At the cloud tops, the differences stem from our choice of $f_{\rm{sed}}$, which we keep fixed at 0.3 in this work, but which can be altered to increase agreement with microphysical models \citep[e.g.,][]{rooney2022,Mang2024}. Future work will explore $f_{\rm{sed}}$ dependences for aggregates (Lodge and Moran et al., in prep).

Overall, our model has decreased cloud mass at the bottom of the cloud but increased mass at the top of the cloud compared to the spherical case of \citet{ohno2020}. On the other hand, our cloud mass is always less than that of the aggregate cloud particle cases in \citet{ohno2020}. Despite different model choices, all of these values are within an order of magnitude or so 
of each other, well within typical variances between microphysical and parametrized cloud models for spherical particles \citep[e.g.,][]{rooney2022}. 

\begin{figure*}
{\includegraphics[width=0.99\textwidth]{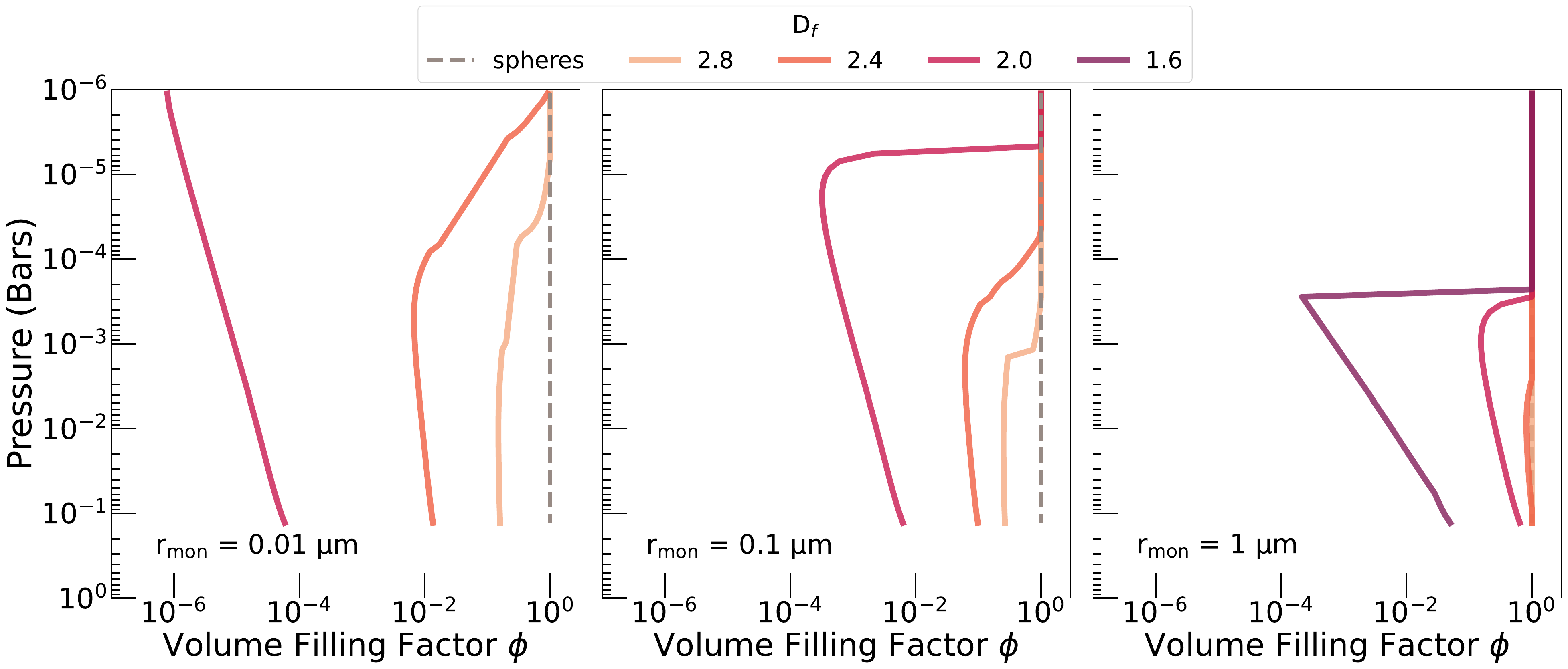}}
{\includegraphics[width=0.99\textwidth]{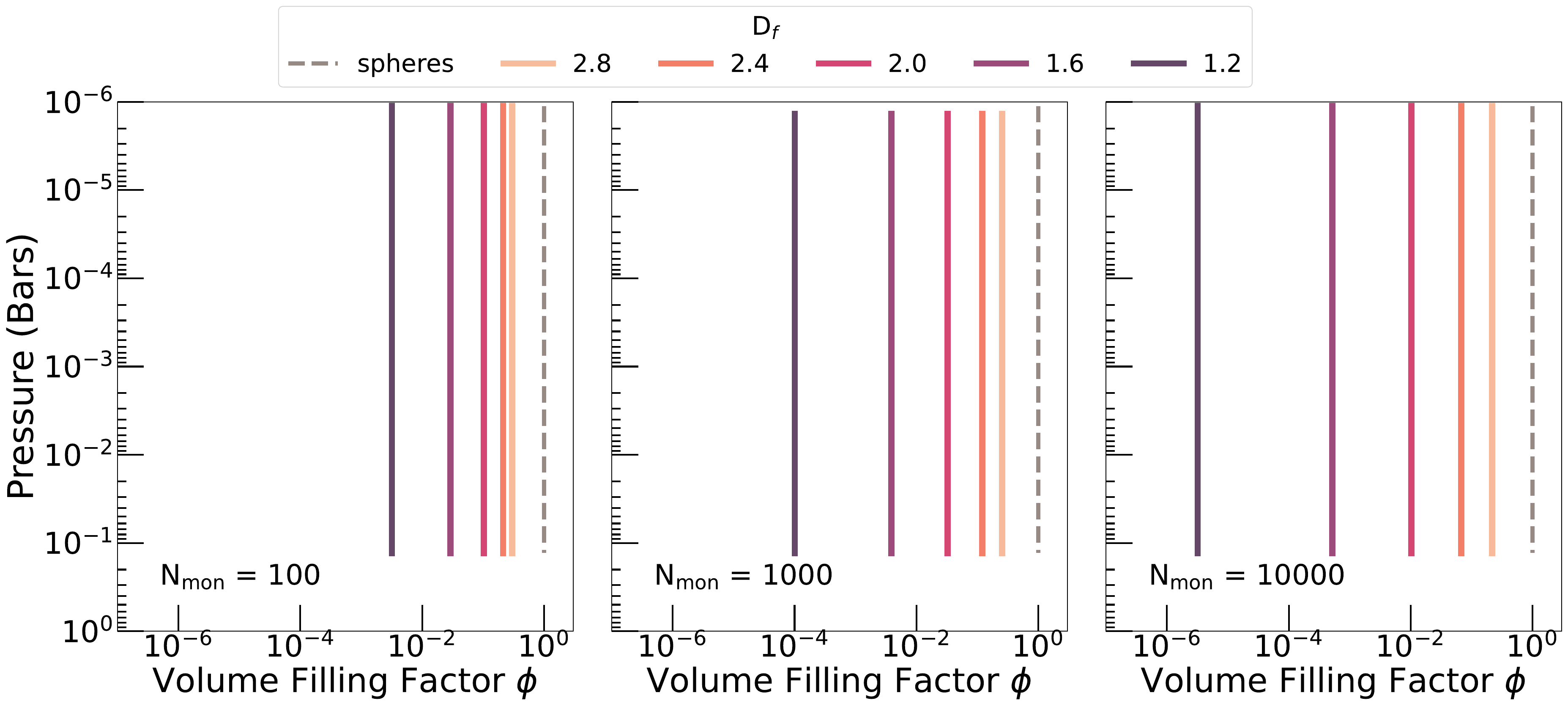}}
\caption{Top: volume filling factor for fixed monomer sizes of $r_{\rm{mon}}$ = 0.01, 0.1, and 1 ${\rm{\mu}}$m as a function of pressure for our model GJ~1214~b atmosphere. Bottom: the same as top, but for fixed $N_{\rm{mon}}$ = 100, 1,000, and 10,000 monomers. As in Figure \ref{fig:particlesizes}, the $r_{\mathrm{mon}} =0.01, D_{\rm{f}} = 2$ particles in the plot in the top left would likely compress to smaller values than those shown here in a real atmosphere. Consequently such particles would have a larger filling factor than shown.}
\label{fig:ff}
\end{figure*}

As detailed in Section \ref{sec:dynamics}, it is possible to generate atmospheric profiles in which no fractal aggregate of a particular type can form. Indeed, we find that for fixed $r_{\rm{mon}}$, we are unable to find any size of aggregate radius $R_{\rm{agg}}$ that can balance the convective velocity $w_{\rm{convect}}$ given our prescribed $K_{\rm{zz}}$ profile for monomers of $r_{\rm{mon}}$ = 0.01 or 0.1 $\mu$m for fractal dimensions smaller than $D_{\rm{f}}$ = 2.0. 
This is because, for $D_{\rm f}\leq2$, the terminal velocity is independent of the aggregate size and equivalent to that of a constituting monomer. When we increase the size of the monomer to $r_{\rm{mon}}$ = 1 $\mu$m, we are able to find stable particle radii for $D_{\rm{f}}$ = 1.6, but not the laciest $D_{\rm{f}} = 1.2$ case. This can be intuitively seen in the top panel of Figure \ref{fig:vfall}, where the lacier $D_{\rm{f}}$ curves for each fixed $r_{\rm{mon}}$ fractal do not cross the dotted line indicating the convective velocity at the pressure of the cloud base.

We show in Figure \ref{fig:particlesizes} the mean particle radii for each of our aggregate models compared to that of the spherical particle case. Here, the mean particle radii are expressed as $r_{\rm{agg},m}$, i.e., in terms of the equivalent mass compact sphere.  As expected, all aggregate particles grow to larger sizes throughout the cloud layer, with particles made of smaller monomers growing correspondingly larger. Comparing the fixed $r_{\rm{mon}}$ to fixed $N_{\rm{mon}}$ implementations, we see that the particle sizes vary less for our set number monomer compared to when the monomer size is set. This result stems from how $r_{\rm{mon}}$ is calculated from $N_{\rm{mon}}$, and the inherently smaller range explored in our fixed $N_{\rm{mon}}$ grid points, which keeps a relatively small difference between the equivalent compact spherical particle radius and the monomer radius.

\begin{figure*}
\centering
{\includegraphics[width=0.99\textwidth]{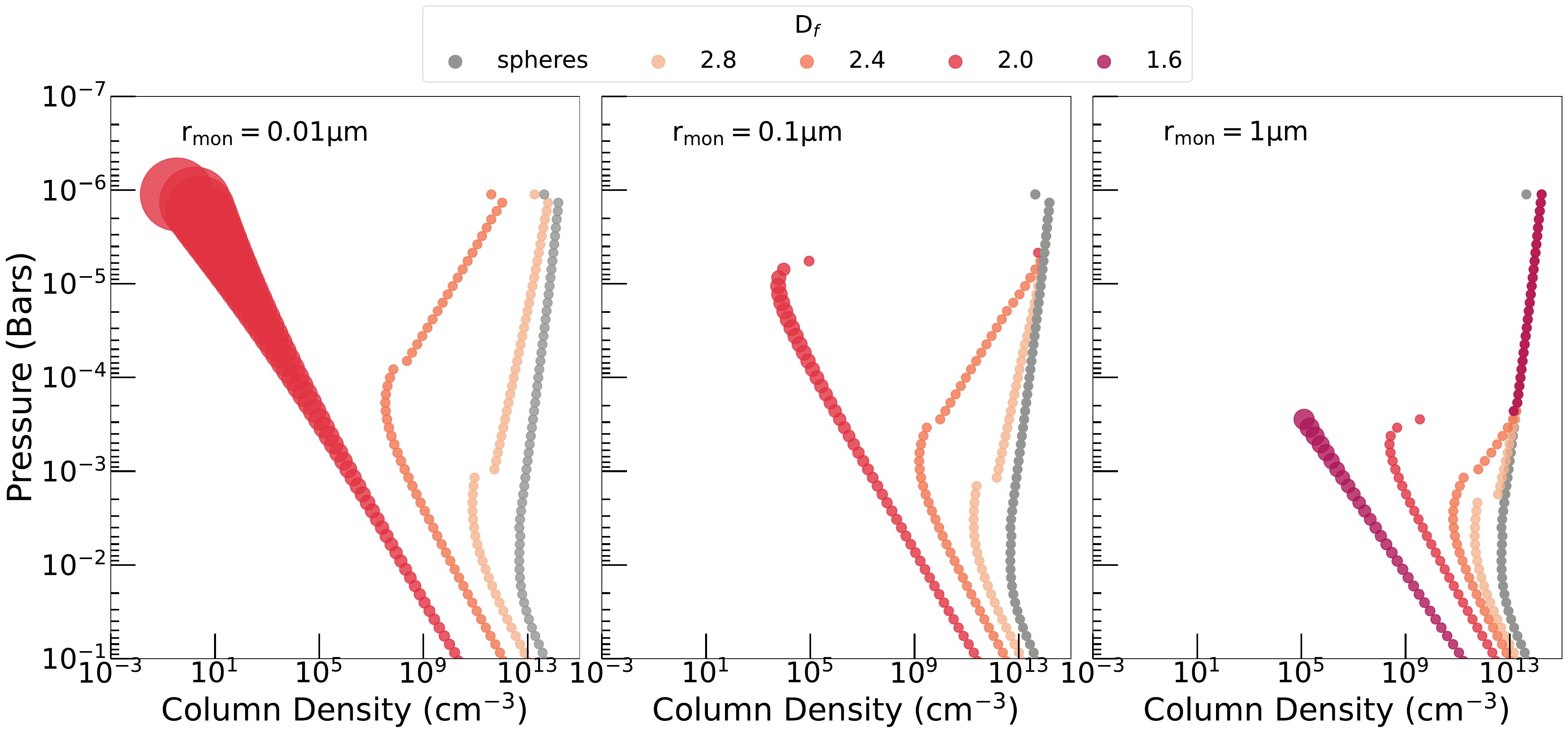}}
{\includegraphics[width=0.99\textwidth]{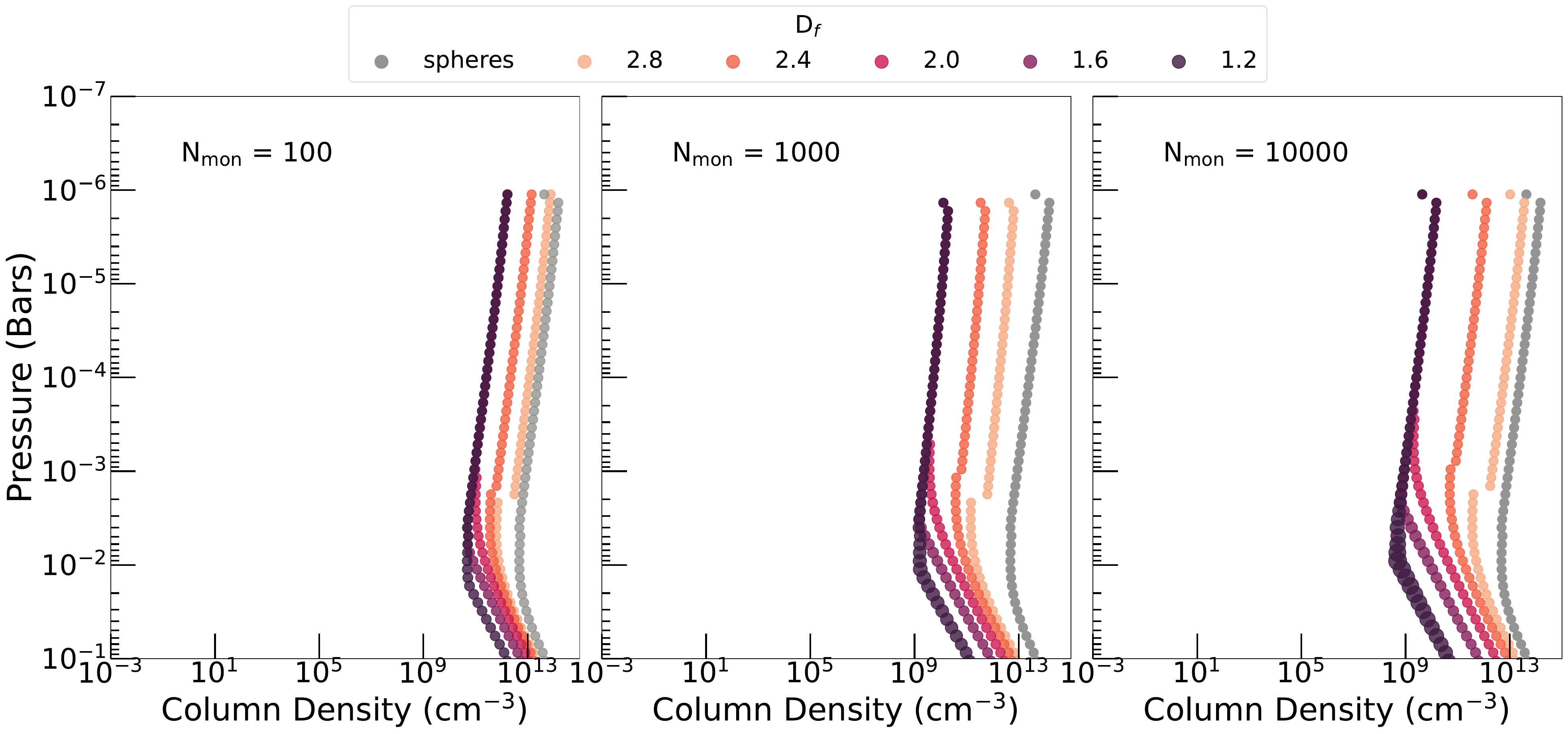}}
\caption{Top: the number of particles as a function of pressure for fixed monomer sizes of $r_{\rm{mon}}$ = 0.01, 0.1, and 1 ${\rm{\mu}}$m for our model GJ~1214~b-like atmosphere. Bottom: the same as top, but for fixed $N_{\rm{mon}}$ = 100, 1,000, and 10,000 monomers. 
The marker sizes are scaled by the equivalent mass compact radius, $r_{\rm{agg},m}$. As in Figure \ref{fig:particlesizes}, the $r_{\mathrm{mon}} =0.01, D_{\rm{f}} = 2$ particles in the plot in the top left would likely compress to smaller than the values here in a real atmosphere.}
\label{fig:dndrmax}
\end{figure*}

The smallest set of monomer sizes, $r_{\rm{mon}}$ = 0.01 $\mu$m, reaches larger particle sizes at higher altitudes, which falls off as the monomer size is increased. Similar aggregate radii of a few microns at the cloud base, as well as similar trends across changing monomer sizes, are found by \citet{ohno2020} for our overlapping $D_{\rm{f}}$ = 2.0 case, though their particle radii are effectively isobaric from $\sim$1$\times$10$^{-3}$ bar to the top of the atmosphere, while ours decrease with height. Note that \citet{ohno2020} adopted the moment method with the monodisperse size distribution, which tends to produce isobaric particle sizes. Once the actual width of the size distribution is taken into account, microphysical models might also produce a size decreasing with height, as smaller aggregates would selectively ascend to the upper atmosphere owing to their slower terminal velocities. Additionally, because of our condition that particles smaller than the prescribed $r_{\rm{mon}}$ radii become spheres, our two larger monomer size cases eventually revert to spheres at the top of the atmosphere. In all cases, we of course produce a wider array of aggregate sizes given our additional $D_{\rm{f}}$ values. 

Note also the red dash-dot line in Figure \ref{fig:particlesizes}, the compression radius calculated for this atmosphere by \citet{ohno2020} (their Equation 18). This is the size at which gas-drag would force a $D_{\rm{f}}=2.0$ aggregate of this size\footnote{The compression radius in Figure 3 of \citet{ohno2020} is a radius of gyration, so we have converted it to a compact mass-equivalent radius ($r_{\rm{agg,m}}$) accordingly for each value of $r_{\rm{mon}}$.} to compress itself into a more compact form. Our $r_{\rm{mon}}$ = 0.01 $\mu$m, $D_{\rm{f}}$ = 2.0 run generates particle radii that exceed this value by an order of magnitude even in the lower atmosphere and nearly two orders of magnitude in the upper atmosphere. This result suggests that these particle shapes would not actually persist here, and in reality they might compress. Given the low masses, high porosities, and thus low opacities of such particles at these pressures, however, the observational impact of this unphysical result is relatively small (see Section \ref{sec:spectra}). Our $r_{\rm{mon}}$ = 0.1 $\mu$m, $D_{\rm{f}}$ = 2.0 run generates some particle radii a factor of $\sim$4 larger than the compression radius, which suggests mild compression is possible but not to the point of strongly impacting the spectrum. The $D_{\rm{f}}$ = 2.0 fractals for fixed $N_{\rm{mon}}$ do not hit the compression radius at any point -- this is perhaps another reason why this option is safer to use for any given atmosphere.


Figure \ref{fig:ff} shows the volume filling factor, $\phi$, for each of our cloudy models. The filling factor is defined as:
\begin{equation}
    \phi = N_{\rm{mon}} \left(\frac{r_{\rm{mon}}}{R_{\rm{agg}}} \right)^3 .
\end{equation}

\noindent Thus, Figure \ref{fig:ff} for fixed $r_{\rm{mon}}$ directly follows inversely from Figure \ref{fig:particlesizes} since we conserve condensate mass at each atmospheric layer. Larger aggregates are more porous than smaller aggregates given a finite amount of cloud mass available. This trend follows the general behavior from \citet{ohno2020} that more porous aggregates form from smaller set monomer radii. However, our fixed $r_{\rm{mon}}$ particles reach filling factors nearly 3 orders of magnitude smaller than that of \citet{ohno2020}. This somewhat balances out the fact that our cloud mass is underestimated compared to theirs as seen in our Figure \ref{fig:kzz} and their Figure 3. Finally, the physical implausibility of our $r_{\rm{mon}}$ = 0.01 $\mu$m, $D_{\rm{f}}$ = 2.0 run is especially strongly emphasized in filling factor space. This model creates aggregate radii significantly larger than the compression radius of $R_{\rm{agg}
}$ $\sim$ 30 $\mu$m ($r_{\rm{agg,m}}$ $\sim$ 2.5 $\mu$m) at pressures less than 1$\times$10$^{-3}$ bar, which corresponds to a filling factor $\phi$ that approaches 10$^{-6}$ at the top of the atmosphere. As shown in the lower panel of Figure~\ref{fig:ff}, the fixed $N_{\rm{mon}}$ cases avoid implausible filling factors, with the same fractal number cases ($D_{\rm{f}}$ = 2.0) behaving very similarly to the aggregates found by \citet{ohno2020}. Only the very laciest aggregates ($D_{\rm{f}}$ = 1.2) reach very low filling factors, as would be expected for such lacy, quasi-linear structures.

\begin{figure*}
{\includegraphics[width=0.99\textwidth]{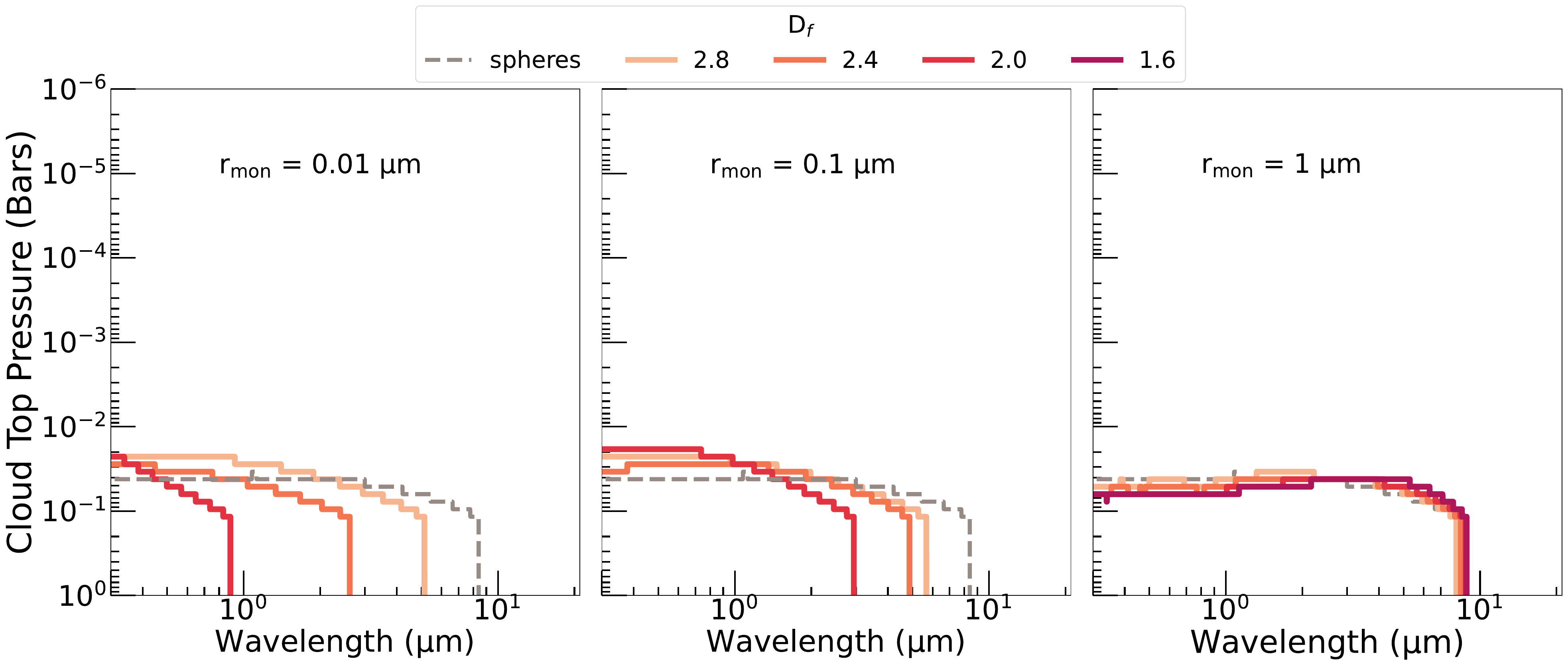}}
{\includegraphics[width=0.99\textwidth]{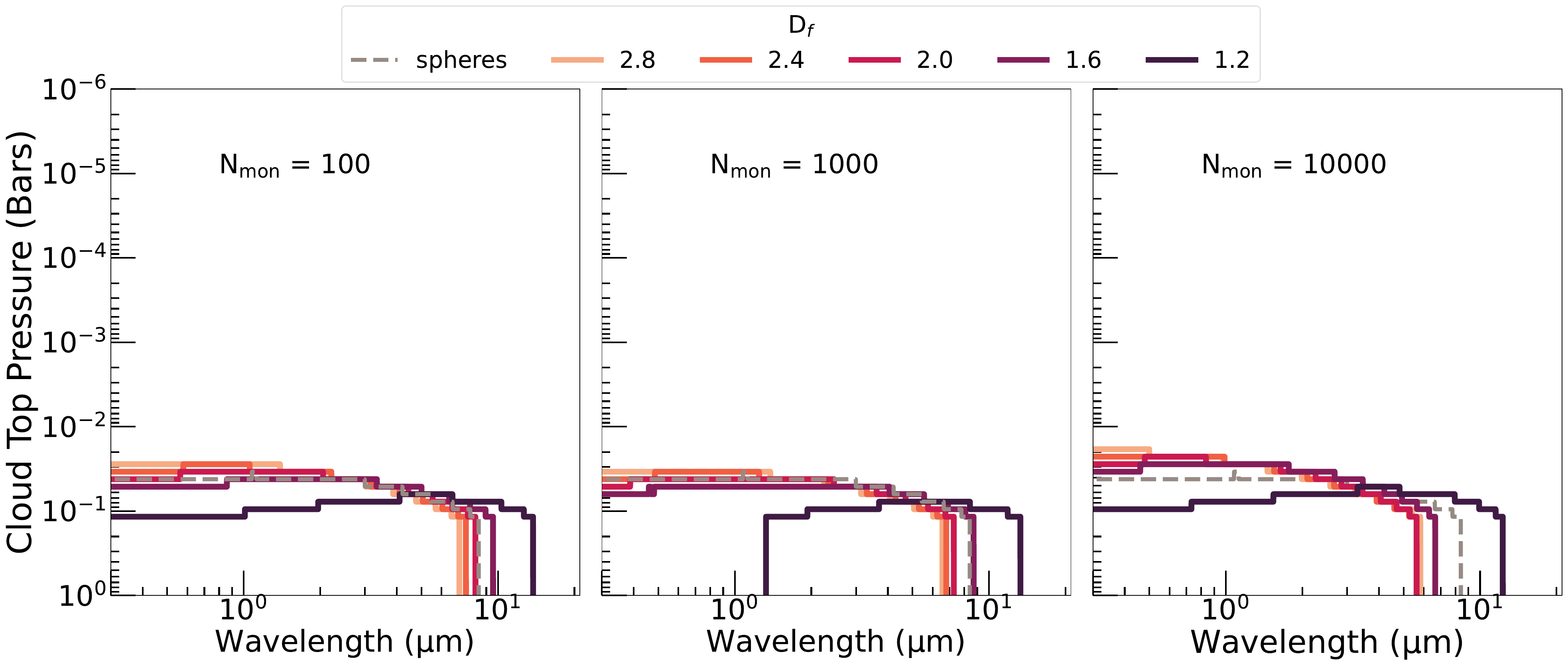}}
\caption{Top: the cloud top pressure (i.e., where $\tau = 1$) for fixed monomer sizes of $r_{\rm{mon}}$ = 0.01, 0.1, and 1 ${\rm{\mu}}$m as a function of pressure for our model GJ~1214~b-like atmosphere. Bottom: the same as top, but for fixed $N_{\rm{mon}}$ = 100, 1,000, and 10,000 monomers.}
\label{fig:tau}
\end{figure*}

Figure \ref{fig:dndrmax} illustrates the number of particles at each pressure layer, i.e., the column density per pressure interval. This visualization provides further intuition for how our cloud mass conservation requirement results in various atmospheric particle size distributions under differing fractal assumptions. The laciest particles have the least number of particles per layer, because fewer particles -- of larger radii -- are needed to make up the difference in mass as the particles become increasingly porous. In a sense, therefore, the column density is just a rescaling of the volume filling factor $\phi$ with the particle size, which can be readily observed by the similarity of the curve shapes in Figures \ref{fig:particlesizes}, \ref{fig:ff} and \ref{fig:dndrmax}.

Again, the fixed $N_{\rm{mon}}$ implementation varies much less than the fixed $r_{\rm{mon}}$ version since the discrepancy in size between monomer radii and aggregate radii is smaller for the values of fixed $N_{\rm{mon}}$ shown. In general, fixed $N_{\rm{mon}}$ behaves more stably compared to the fixed $r_{\rm{mon}}$ versions, which is demonstrated here.

Finally, our Figure \ref{fig:tau} compares the wavelength-dependent cloud top pressure for each of our model runs. As is typical, we define the cloud top pressure as the point where the cumulative optical depth, $\tau$, is equal to or approaches unity. For each of our runs, $\tau$ approaches 1 somewhere between 0.1 and 0.01 bar. For fixed $r_{\rm{mon}}$, lacier (i.e., more porous) aggregates drop off in optical depth first as the wavelength increases following the well-known behavior for such particles \citep[e.g.,][]{sanaz2024}. As the size discrepancy between the monomer radius and the aggregate radius narrows, the optical depth across cases approaches that of a sphere.

For fixed $N_{\rm{mon}}$ the laciest $D_{\rm{f}}$ = 1.2 runs are noticeably less opaque than the others, even at the shortest, visible wavelengths. This decreased opacity is purely an outcome of the optics of such lacy aggregates, as seen in Figure~\ref{fig:opacity}. 
The cloud top still contributes noticeably to the spectrum (see Section \ref{sec:spectra}), but the pressures at which this occurs is markedly deeper compared to the other more compact aggregates. Additionally, this very lacy aggregate run also maintains an optical depth of unity to longer wavelengths compared to the more compact particles -- beyond 10 $\mu$m. This behavior results from the combination of larger aggregate radii (see Figure \ref{fig:particlesizes}) combined with relatively non-porous particles (Figure \ref{fig:ff}) at relatively high abundances (Figure \ref{fig:dndrmax}) in the lower ($\sim$1-10$^{-2}$ bar) atmosphere. 

Our cloud top pressures are markedly deeper compared to the cloud tops observed by \citet{ohno2020}. While all of our cloud tops lie around $\sim
$10$^{-1}$ to 10$^{-2}$  bar, \citet{ohno2020}'s cloud tops vary from 9$\times$10$^{-3}$ to 10$^{-5}$ bar in the visible depending on the monomer size, and around 10$^{-2}$ bar in the infrared. The same wavelength behavior with monomer size occurs in both our cloud tops, just at different pressures. Again, this is driven by the differing cloud mass profiles of our two cloud modeling schemes. 

\begin{figure*}
\centering
{\includegraphics[width=0.9\textwidth]{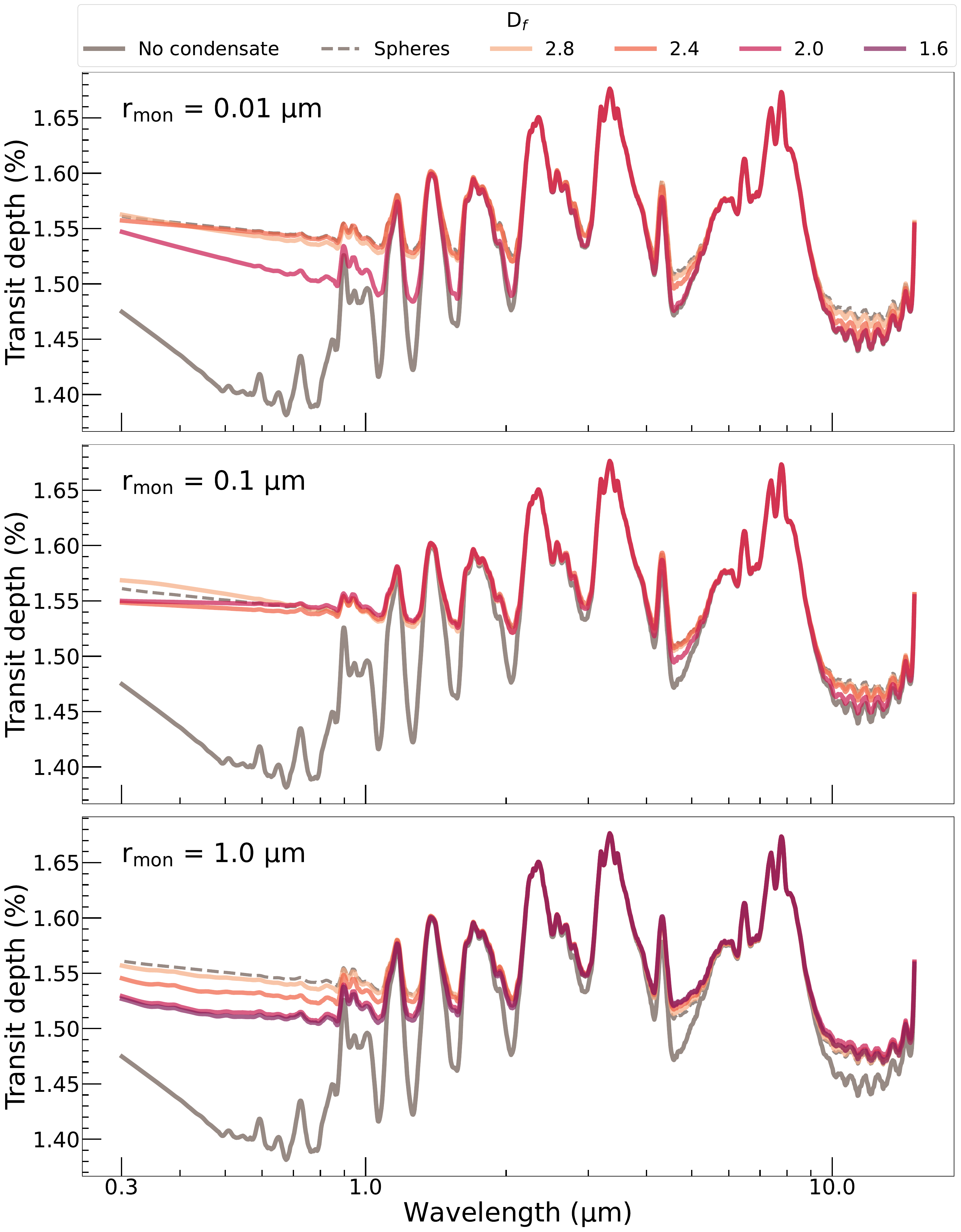}}
\caption{Top, center, bottom: Simulated transmission spectra for aggregate KCl clouds in a 100$\times$ solar metallicity atmosphere on a GJ~1214~b-like planet using fixed $r_{\rm{mon}}$ = 0.01, 0.1, and 1 ${\rm{\mu}}$m sized monomers, respectively, for fractal dimensions of $D_{\rm{f}}$ = 1.2, 1.6, 2.0, 2.4, and 2.8 along with spherical cloud particles. The change in opacity at visible-NIR wavelengths is highly dependent on the size of monomer. For KCl clouds, lacier aggregates generally decrease the opacity overall, but this is sensitive to the exact location of the cloud deck.}
\label{fig:spectra_rmon}
\end{figure*}

\begin{figure*}
\centering
{\includegraphics[width=0.9\textwidth]{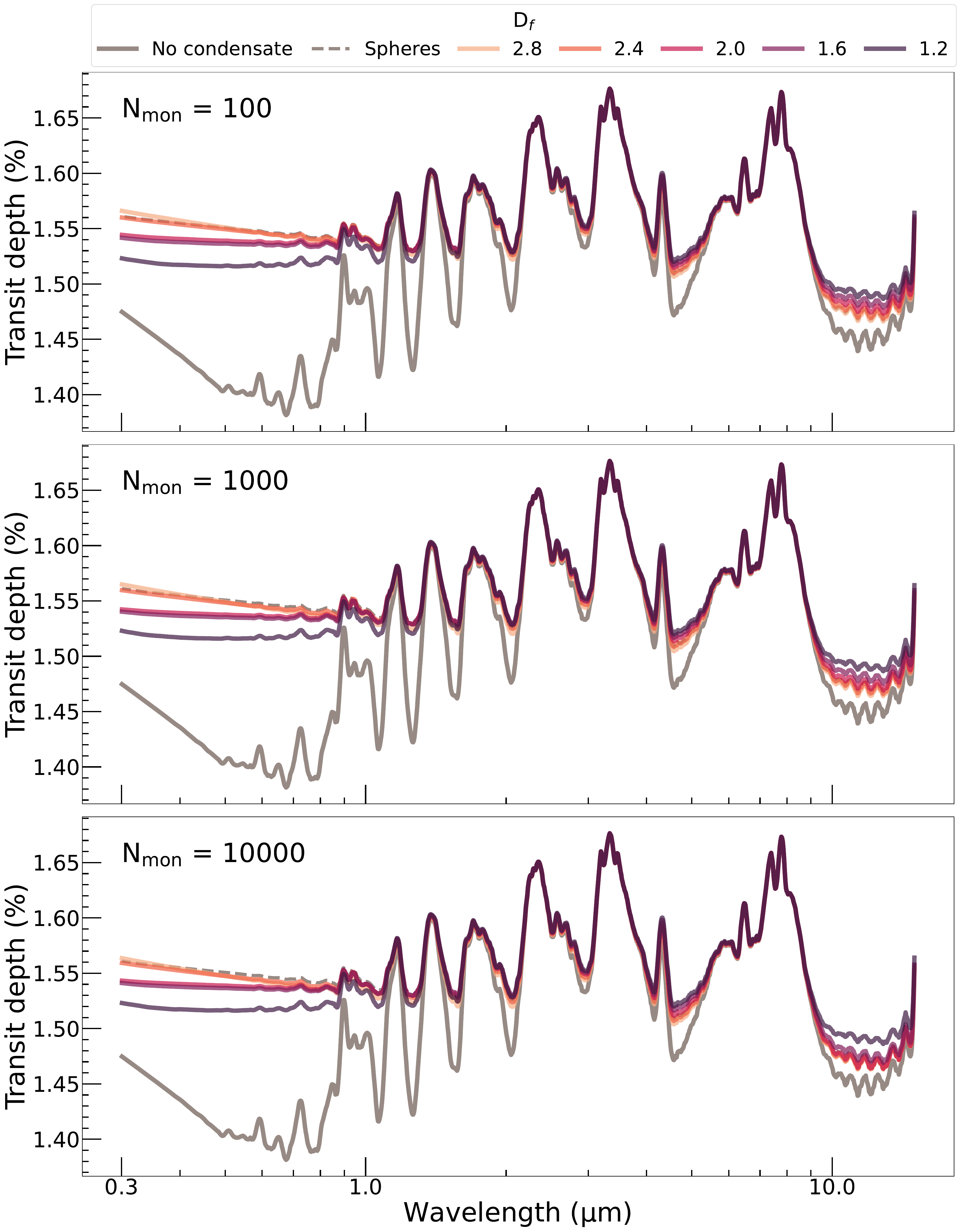}}
\caption{Top, center, bottom: Simulated transmission spectra for aggregate KCl clouds in a 100$\times$ solar metallicity atmosphere on a GJ~1214~b-like planet using fixed $N_{\rm{mon}}$ = 100, 1,000, and 10,000 monomers, respectively, for fractal dimensions of $D_{\rm{f}}$ = 1.2, 1.6, 2.0, 2.4, and 2.8 along with spherical cloud particles. Changing the number of monomers in this formulation hardly changes the spectra. For KCl clouds, lacier aggregates increase the opacity at longer wavelengths and decrease the opacity at shorter wavelengths compared to spheres.}
\label{fig:spectra_Nmon}
\end{figure*}



\subsection{Simulated Transmission Spectra with Aggregate Clouds}\label{sec:spectra}

We demonstrate the observable effects of our aggregate clouds on the transmission spectra of our GJ~1214~b-like planet's atmosphere. We generate these transmission spectra using {\picaso}'s radiative transfer module with the V3 opacity database from 0.3 to 15 $\mu$m resampled at R=15,000 from a line-by-line calculation performed at R$\sim$1e6, available on Zenodo\footnote{\href{https://zenodo.org/records/14861730}{PICASO Opacities V3}}. This database contains \ce{C2H2}, \ce{CH4}, CO, \ce{CO2}, \ce{H2-H2}, \ce{H2O}, K, Na, \ce{H2S}, OCS, and \ce{SO2}, which are the most important molecular absorbers from 0.3 to 15 $\mu$m in a 100$\times$ solar atmosphere.  The spectra shown are rebinned to a resolution of R=300. Our simulated spectra therefore cover the wavelength range accessible to Hubble and JWST's instruments conducive for atmospheric observations in transmission.

As shown in Figures \ref{fig:spectra_rmon} and \ref{fig:spectra_Nmon} for fixed $r_{\rm{mon}}$ and fixed $N_{\rm{mon}}$ respectively, the lacier aggregate clouds of KCl always decrease the strength of the scattering in the visible compared to more compact aggregates or spheres. The presence of any cloud mutes gaseous spectral features throughout the near-infrared because of the additional source of opacity, with lacier aggregates muting the gaseous features less as they contribute less opacity. 

Aggregates behave differently in the mid-infrared depending on whether the fixed monomer number or fixed monomer radius is used. For fixed monomer radii, the lacier aggregates are less absorptive compared to spheres or compact particles, while for fixed monomer number, the laciest aggregates are actually more absorbing compared to compact particles or spheres. The cloud optical depth plots shown in Figure \ref{fig:tau} clearly show this, which arises from combination of the large aggregate radii made of fairly non-porous particles in moderate abundance under fixed $N_{\rm{mon}}$. In addition, because the fixed monomer number runs can find a solution for the laciest aggregates with $D_{\rm{f}}$ = 1.2, the differences in the scattering slopes in the visible are larger compared to the fixed monomer radii cases which have stable solutions only for more compact aggregates.

Because we use KCl for our cloud composition, the spectra are simply muted across all wavelengths shown, as KCl does not have distinct spectral features until the far-infrared, past 20 $\mu$m, which is beyond the range of JWST for transiting exoplanet studies. Future work will explore other compositions that have significant wavelength dependence to examine how this interacts with aggregate opacity effects (Lodge and Moran et al. in prep). 

Finally, we compare our simulated transmission spectra to those of \citet{ohno2020} and \citet{Adams2019}. Compared to \citet{ohno2020}, our $D_{\rm{f}} = 2$ cases with set monomer sizes -- the models directly comparable to theirs -- reproduce their transmission spectra very well. We capture the steeper scattering slope present with $r_{\rm{mon}}$ = 0.01 $\mu$m and the flatter slope through the visible under large monomer radii assumptions. The scale of muting throughout the near- and mid-infrared is reproduced as well, on the order of a few hundred parts-per-million (ppm). Because we explore additional fractal dimensions, we also see a wider range of behavior for each set of models. For example, we produce greater variances in scattering slope differences, on the order of 500 ppm, whereas \citet{ohno2020} only saw up to $\sim$400 ppm differences. 

The aggregate haze models of \citet{Adams2019} produced extremely flat spectra across the range observable by Hubble and JWST, from 0.3 to 30 $\mu$m. As discussed by \citet{ohno2020}, this primarily stems from their choice of fractal dimension for large numbers of monomers, where $D_{\rm{f}}$ was assumed to be 2.4. In addition, since they used photochemical hazes rather than clouds, they produced large particle masses very high in the atmosphere, which results in strong spectral flattening. Our higher fractal dimension cases of 2.4 and 2.8 are also flatter, though the exact slope in the spectrum is highly dependent on the monomer size, whether set directly or calculated from the number of monomers when $N_{\rm{mon}}$ is set. 
We also note that \citet{Adams2019} adopted the refractive indices of soot and Titan tholin \citep{khare1984} that are much more absorptive than the KCl assumed here, which also efficiently flattens the spectrum \citep[e.g.,][]{ohno2020b,2023ApJ...951..117S}.
Still, we demonstrate that our model is able to more flexibly sample a wider range of aggregate particle shapes compared to previous microphysical models while maintaining a level of physical plausibility for cloud formation.


\subsection{Why are some fluffy aggregate clouds in {\virga} transparent?}
As seen in Figures \ref{fig:spectra_rmon} and \ref{fig:spectra_Nmon}, the impact of aggregate morphology is relatively modest, and some models with different $D_{\rm f}$ nearly overlap each other. 
This may seem odd, as fluffy particles have large surface area per mass and remain in the atmosphere for a long time thanks to slow settling velocity.

We here analytically demonstrate that the counterintuitive trend comes from the intrinsic formulation of the original \citet{ackerman2001} {\eddysed} framework.
The {\eddysed} model first calculates the vertical distribution of cloud mass mixing ratio for a given set of $f_{\rm sed}$ and $K_{\rm zz}$ on the basis of Equation \eqref{Eq:eddysed}.
Inserting $w_{\rm *}=K_{\rm zz}/H$ and $q_{\rm t}\approx q_{\rm c}$, which is valid well above the cloud base where most vapor has already condensed to particles, one can analytically solve Equation \eqref{Eq:eddysed} to obtain the cloud mass mixing ratio as 
\begin{equation}
    q_{\rm c}=q_{\rm c0}\left( \frac{P}{P_{\rm 0}}\right)^{f_{\rm sed}},
\end{equation}
where we have used hydrostatic balance to adopt atmospheric pressure as our vertical coordinate.
Importantly, the cloud mass profile is solely determined by $f_{\rm sed}$ and independent of any aerosol properties such as particle size and fractal dimension.
Thus, no matter how fluffy the aerosol is, clouds have the same vertical abundance profile for a given $f_{\rm sed}$.
This contrasts with microphysical models that yield more vertically extended clouds for fluffy particles \citep{ohno2020}. 

On the other hand, optical properties do depend on the aggregate morphology.
Under the approximation that incident light is only singly scattered by monomers within the aggregate, which approximates the scattering cross section as (Appendix A of \citealt{ohno2021haze}, see also \citealt{Berry&Percival86}):
\begin{equation}\label{eq:Csca_agg}
    C_{\rm sca, agg} \approx N_{\rm mon}C_{\rm sca, mon}\frac{2N_{\rm mon}^{(D_{\rm f}-2)/D_{\rm f}}}{(D_{\rm f}-1)(2-D_{\rm f})}\left( \frac{b \lambda}{4\pi r_{\rm mon}}\right)^{2},
\end{equation}
where $b=\sqrt{(D_{\rm f}+1)D_{\rm f}}$ \citep{Berry&Percival86}, and we have picked the formula applicable to $D_{\rm f}>2$ for illustrative purposes.
From the definition of $f_{\rm sed}\equiv v_{\rm fall}/w_{\rm *}=Hv_{\rm fall}/K_{\rm zz}$, for the kinetic regime of gas drag, we have a relation between $f_{\rm sed}$ and other aggregate properties as
\begin{equation}\label{eq:fsed_vfall}
    f_{\rm sed}\frac{K_{\rm zz}}{H}=\frac{2g\rho_{\rm m}}{3v_{\rm th}\rho_{\rm a}}r_{\rm mon}N_{\rm mon}^{(D_{\rm f}-2)/D_{\rm f}},
\end{equation}
where we have used Equation \eqref{eq:v_agg_sanaz2024} and $r=r_{\rm mon}N_{\rm mon}^{1/3}$.
Combining Equation \eqref{eq:Csca_agg} and \eqref{eq:fsed_vfall}, we obtain the mass scattering opacity as
\begin{align}
    \begin{split}
        \kappa_{\rm sca,agg}&\equiv\frac{C_{\rm sca,agg}}{M_{\rm agg}}\equiv\frac{C_{\rm sca,agg}}{N_{\rm mon}m_{\rm mon}} \\
        &\approx \zeta\frac{C_{\rm sca,mon}}{m_{\rm mon}r_{\rm mon}}\left( \frac{\lambda}{4\pi r_{\rm mon}}\right)^2
    \end{split}
\end{align}
where
\begin{equation}
    \zeta=\frac{3v_{\rm th}\rho_{\rm a}f_{\rm sed}K_{\rm zz}}{g\rho_{\rm m}H}\frac{D_{\rm f}(D_{\rm f}+1)}{(D_{\rm f}-1)(2-D_{\rm f})}
\end{equation}
Recalling that the scattering cross section of monomers follows $C_{\rm sca,mon}\propto r_{\rm mon}^6$ in the Rayleigh regime, and that $m_{\rm mon}=\frac{4}{3}\pi r_{\rm mon}^3$:
\begin{align}
    \begin{split}
    \kappa_{\rm sca,agg}&\propto \zeta \frac{r_{\rm mon}^6} {r_{\rm mon}^3 r_{\rm mon}}\left( \frac{\lambda}{4\pi r_{\rm mon}}\right)^2 \\
    \kappa_{\rm sca,agg}&\propto \zeta \lambda ^2
    \end{split}
\end{align}
We therefore find that the mass scattering opacity $\kappa_{\rm sca,agg}$ turns out to be independent of monomer size and only weakly depends on the fractal dimension
for $D_{\rm{f}}$ $\gtrapprox$ 2.2 if the terminal velocity is the same. However, when $D_{\rm{f}}$ $\lessapprox$ 2.2, the $\zeta$ term drops sharply, and thus so does the corresponding mass scattering opacity, resulting in more transparent particles. 

For the fixed $r_{\rm{mon}}$ explored here, only the $r_{\rm{mon}}~=~0.01$ $\mu$m monomers are in the Rayleigh regime across the full wavelength range shown, and thus this result is demonstrated most strongly in the top panel of the fixed $r_{\rm{mon}}$ models, where the $D_{\rm{f}}$ = 2 model is notably more transparent with lower transit depth. For fixed $N_{\rm{mon}}$, monomers are in the Rayleigh regime in the upper atmosphere, i.e., less than 10$^{-5}$ bar and lower pressures, for all models.  The laciest cases extend monomers in the Rayleigh regime down to pressures of 0.1 bar for $N_{\rm{mon}}$ = 10,000 but only down to 10$^{-4}$ bar for $N_{\rm{mon}}$ = 100. Thus, the impact of the more transparent aggregates is more modulated for the fixed $N_{\rm{mon}}$ cases -- the mass mixing ratio is already quite low in regions of atmosphere where this approximation holds, as shown in the right panel of Figure~\ref{fig:kzz}.

\section{Discussion and Conclusion} \label{sec:conclusion}

We have introduced a new version of the {\virga} cloud model --released as Version 2.0-- which is capable of accounting for both the dynamical and optical consequences of fractal aggregate cloud particles in a self-consistent way. Our model is flexible, fast, and ideal for parameter space sweeps given an atmospheric spectrum that has evidence for muting or enhanced scattering by clouds, especially when a combination of both small and large particle effects are observed. 

While our model is highly flexible and can reproduce many of the broad aggregate cloud behaviors observed in true microphysical models, it still lacks certain features. First, we caution users that setting the monomer radii by fixed $r_{\rm{mon}}$ can often result in no solution for a stable aggregate radius, as these particles are extremely fluffy and resist sedimentation. Users should always inspect the physical parameters of their {\virga} outputs rather than heedlessly using the spectra these models produce. We enable such inspections by providing tutorials in the online documentation of our code which examines several diagnostic properties, as we have shown in this work. We also note that the fixed $N_{\rm{mon}}$ implementation does not suffer from these issues and can frequently provide much more stable and reasonable solutions that still reasonably approximate aggregate growth.

As with {\virga} generally, whether particles are spheres or fractal aggregates, we ignore microphysical processes, including but not limited to nucleation, coagulation, compression, or break-up. As many of these processes are composition-dependent and the result of any given species' particular material properties \citep[e.g.,][]{gaobenneke2018}, our model likely frequently overestimates condensate masses for certain condensible species. Our particles are also all of the same fractal dimension with self-similar monomers, as in \citet{ohno2020}, while in reality, aggregates can form multiple different shapes and sizes throughout the atmospheric column \citep{Adams2019,OhnoTanaka21}. In addition, we have modeled only one cloud composition -- KCl -- when mixed composition particles are probable and would also influence particle morphology \citep{Samra2020,Samra2022,kiefer2024}. 

{\virga} also assumes a lognormal particle size distribution, while other options, such as a Hansen distribution \citep{hansen}, gamma distribution \citep[e.g.,][]{christie2022,LeeOhno2025} or a double peaked distribution of coarse and fine-mode particles, may also occur or even be more likely, and frequently form in microphysical model schemes \citep[e.g.,][]{Powell2019}. In particular, how and whether fractal aggregates would form a lognormal size distribution is unknown, and the above other distributions -- or others -- may be even more important for aggregates.

Certain materials are more likely to form aggregate structures compared to others. For example, on Earth, clouds made of water droplets would not be accurately described by fractal aggregate clouds, while soots and secondary organic aerosol \textit{are} frequently observed to be highly fractal \citep{wang2017, adachi2010shapes}. Photochemical hazes made of organics readily coagulate to form long particle chains \citep{Adams2019,yu2020,Yu2021}, and Martian dust appears to easily form aggregates as well \citep[e.g.,][]{CHENCHEN201916}, which could have similar behavior to mineral cloud conglomeration in exoplanets and brown dwarfs \citep{Samra2020,Samra2022}. 

In general, we advise that our fractal aggregate version of {\virga} always be used in conjunction with a microphysical cloud model when interpreting observations of exoplanetary atmospheres. {\virga}'s speed can be utilized to perform a wide parameter space study, and once a cloud morphology and composition is identified, a more limited set of microphysical models can be explored to examine whether the material properties and atmospheric conditions of the planet would support the formation of aggregate aerosol as suggested by {\virga}. The flexibility we have built into the code, with multiple options for defining fractal particles, should enable such comparisons while also broadly capturing a wide range of observable cloud effects in atmospheric spectra.




\nolinenumbers
\begin{acknowledgments}
The authors are grateful to the anonymous referee for their helpful review. S.E.M. is supported by NASA through the NASA Hubble Fellowship grant HST-HF2-51563 awarded by the Space Telescope Science Institute, which is operated by the Association of Universities for Research in Astronomy, Inc., for NASA, under contract NAS5-26555. This project benefited from in-person collaboration at the 2022, 2023, and 2024 Exoplanet Summer Programs at the Other World Laboratory (OWL) at the University of California, Santa Cruz, which are funded by the Heising-Simons Foundation. This work also benefited from discussions at Cloud Zwei Con at Schloss Ringberg, organized by the APEx department of the Max Planck Institute for Astronomy, in particular discussions with Diana Powell, Channon Visscher, Michiel Min, and Duncan Christie.
This work benefited from in-person collaboration between S.E.M and M.G.L funded by the UK Research and Innovation (UKRI) framework under the UK government’s Horizon Europe funding guarantee for an ERC Starter Grant [grant number EP/Y006313/1].
M.G.L acknowledges the generous support of the Keith Burgess Scholarship and Frederick Frank Fund.
H.R.W was funded by UKRI under the UK government’s Horizon Europe funding guarantee for an ERC Starter Grant [grant number EP/Y006313/1].  N.E.B. acknowledges support from NASA’S Interdisciplinary Consortia for Astrobiology Research (NNH19ZDA001N-ICAR) under award number 19-ICAR19\_2-0041.
\end{acknowledgments}

\software{
Astropy \citep{astropy,astropy2,astropy2022},
\texttt{CORAL} \citep{Lodge2024}, Matplotlib \citep{matplotlib}, NumPy \citep{numpy, numpynew}, \texttt{Optool} \citep{dominik2021}, \texttt{PICASO} \citep{batalha2019,Mukherjee2023}, pysynphot \citep{STScIDevelopmentTeam2013},
\texttt{PyMieScatt} \citep{piemiescatt},
\texttt{scipy} \citep{scipy},
{\virga} \citep{batalha2020,rooney2022,batalha2025virga}}

\bibliography{arxiv_version}{}
\bibliographystyle{aasjournal}



\end{document}